\documentclass[twocolumn]{aastex631}

\shorttitle{}
\shortauthors{}

\usepackage{float}
\usepackage{amsmath}

\newcommand{\lsi}{\,\raisebox{-0.13cm}{$\stackrel{\textstyle<}
{\textstyle\sim}$}\,}

\begin{document}

\title{Galaxy rotation curves disfavor traditional and self-interacting dark matter halos,\\ preferring a disk component or Einasto function}

\author[0000-0001-5117-2732]{Nicolas Loizeau}
\affiliation{Center for Cosmology and Particle Physics, New York University}

\author{Glennys R. Farrar}
\affiliation{Center for Cosmology and Particle Physics, New York University}

\begin{abstract}
We use the galaxy rotation curves in the SPARC database to compare 9 different dark matter and modified gravity models on an equal footing, paying special attention to the stellar mass-to-light ratios. We compare three non-interacting dark matter models, a self interacting DM (SIDM) model, two hadronically interacting DM (HIDM) models, and three modified Newtonian dynamics type models:  MOND, Radial Acceleration Relation (RAR) and a maximal-disk model.  The models with DM-gas interactions generate a disky component in the dark matter, which significantly improves the fits to the rotation curves compared to all other models except an Einasto halo; the MOND-type models give significantly worse fits. 
\end{abstract}


\section{Introduction} \label{sec:intro}

The study of galactic dynamics raised a missing mass problem \citep{Rubin1980}. Two main models claim to solve it: Modified Newtonian Dynamics (MOND) and cold dark matter ($\Lambda$CDM) \citep{Milgrom1983, Blumenthal1984}. Even though non-interacting CDM models are very successful, they potentially raise new problems such as the ``core-cusp" problem, the explanation of bulge-less galaxies, the missing-satellite problem and the ``too big to fail problem" \citep{Bullock2017}. Proposed solutions include Self-Interacting DM (SIDM) and the indirect effects of ``gastrophysics": baryon interactions that feed back on the DM via gravity to smooth the cusp \citep[][and many others]{Spergel2000, Ren2019, Pontzen2012, Santos2020}.  Such effects could be enhanced by a DM interaction with baryons.  A specific realization of the latter possibility is that  the dark matter particle is an as-yet-undiscovered neutral stable hadron composed of 6 quarks $uuddss$ (sexaquark) \citep{Farrar2017eqq,fDMtoB18,Farrar2020}. The cross-section of such a DM particle with baryons could potentially be large enough that interactions with gas cause DM to locally take on a similar structure to the gas in the galactic disk \citep{Farrar2017ysn}.  The nature of such a disk would depend on the strength of the coupling, but generically the DM disk would be thicker than the gas disk since DM forming the disk has mostly  scattered only once.  Therefore, the constraints of~\citet{Schutz+thinDMdisk18} do not apply. 

Our goal here is to compare these different models on an equal footing using the best available rotation curves, those of the SPARC database \citep{Lelli2016}.  We selected 9 models to test:  two traditional non-interacting dark matter halos -- NFW and pseudo-isothermal -- a SIDM model \citep{Ren2019}, two models including DM-baryon interactions, MOND, the radial acceleration relation (RAR) ansatz \citep{McGaugh2016}, the maximal-baryon model \citep{Swaters2012} and finally the Einasto functional form which was found to give the best overall fit to SPARC rotation curves, of the functions explored in the study of \citet{Li2020}. 

The main limitation in galaxy rotation curve fitting is the uncertainty on the stellar mass-to-light ratios of individual galaxies. In our comparative study, we pay particular attention to the handling of stellar mass-to-light ratios.
Previous studies used free or fixed stellar mass-to-light ratios for rotation curve fitting. 
Using fixed values based on stellar population synthesis models relies too much on the quality of such models, while letting the mass-to-light ratio of individual galaxies be fully free sacrifices constraining power.  E.g., 
gas scaling fits resulted in an unphysical stellar mass-to-light ratio distribution ranging between 0 and 14 $M_\odot/L_\odot$ \citep{Swaters2012, Noordermeer2006}, strongly discrepant from population synthesis predictions and observations in the literature.  Some more recent work \citep[e.g.,][]{Katz2016} restricts $M_\odot/L_\odot$ to some range but with a flat prior and quite broad range.  Here, we constrain the fit such that the distribution of mass-to-light ratios has a physical range, with the mean value and width as determined in \citet{Lelli2016, Swaters2014, Schombert2014, McGaugh2014, Meidt2014, Schombert2018}.

Our analysis reveals a striking improvement in fit quality for the HIDM models which have a disky DM component reflecting the gas disk (especially the physically-based ``interactions-scaling" model) in comparison to standard alternatives.  Therefore, we explore (Sec.~ \ref{sec:HIDMinterp}) what type of HIDM cross section would be needed to account for the rotation curve results, and find the required cross section can be compatible with DM direct detection limits.  

While there is a physical motivation for allowing a HIDM disk, an improved fit with a disk component does not prove that a disk component exists.  To see whether a comparable improvement can be achieved with a spherically symmetric halo, we explore alternate functional forms. \citet{Li2020} found that the Einasto function, with one more parameter than traditional CDM halos such as NFW or pseudo-isothermal, gave the best fit to the ensemble of SPARC rotation curves of the functions they considered.  We find in our analysis that the Einasto function gives a comparably good fit as the HIDM model.  We investigate whether the Einasto fits display any distinctive characteristics such as occupying some subset of the allowed parameter space, but do not uncover any regularities.  Future work is needed to understand whether the Einasto form gives a superior fit to simulated galaxies than traditional CDM halo functions, and whether a DM disk may form in simulations  with only gravitational DM-baryon interactions.  Strategies for observationally distinguishing between spherically symmetric and non-spherical DM distributions are also needed; a new approach is proposed in \citet{Loizeau2021}.

\section{The data and models}
\label{sec:data}

\begin{figure*}
\plotone{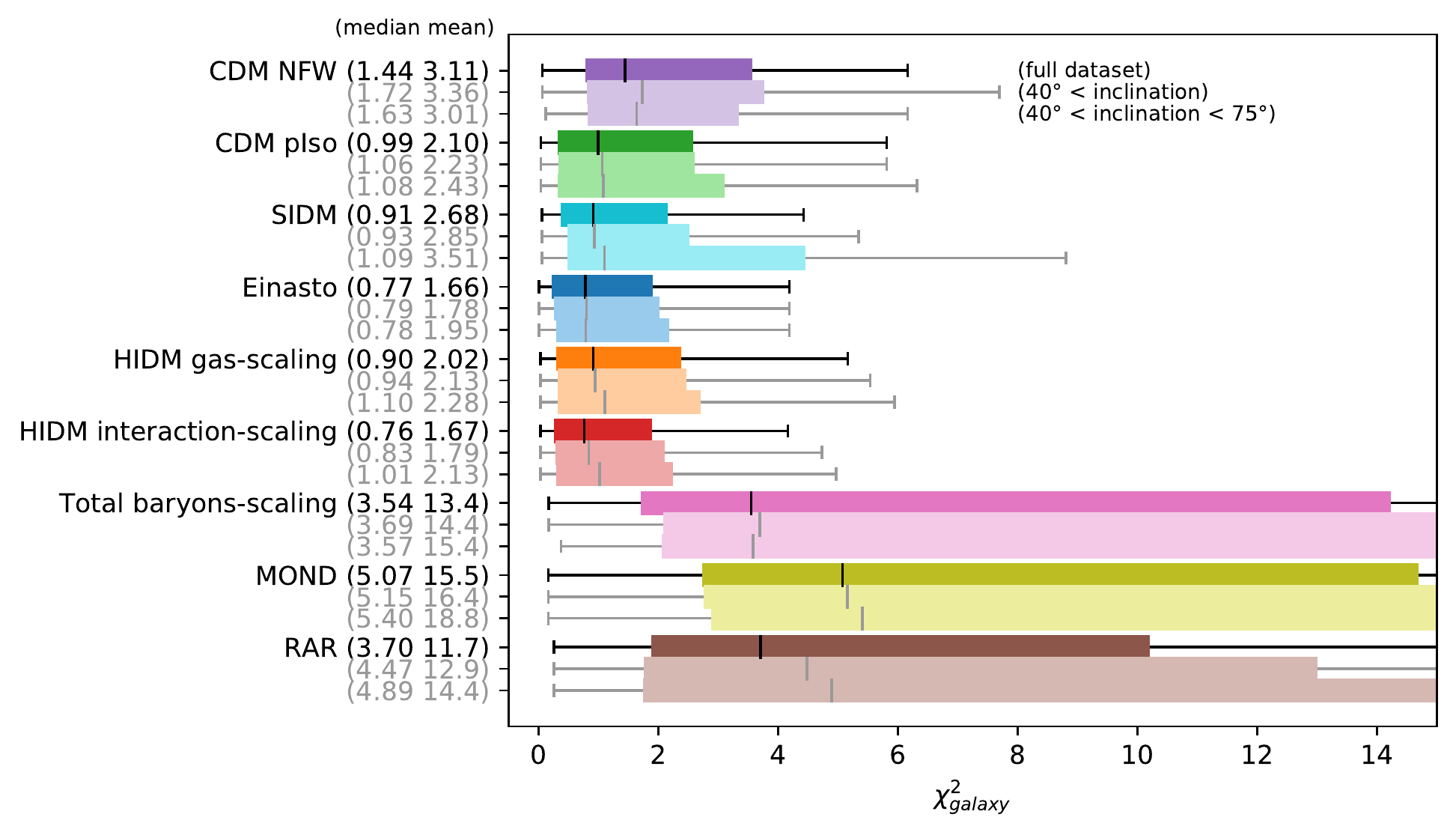}
          \vspace{-0.1in}
\caption{Reduced $\chi^2$'s for the models considered, with the median and mean value in ()'s after the model name. The top line for each model corresponds to the entire dataset of 121 galaxies while the middle line is for the restricted dataset with inclinations greater than $40^\circ$ and the bottom line is the restricted dataset with inclinations between $40^\circ$ and $75^\circ$. The vertical bar is the median of the individual galaxies' reduced $\chi^2$, the box contains the values between the first and third quartiles $Q1$ and $Q3$, and the right (left) whisker extends to the first $\chi^2$ greater (less than) than $Q1 \pm 1.5\, (Q3-Q1) $.  The whiskers on the total-baryons-scaling, MOND and RAR models extend to $\chi^2 = 24, 31, 21$, respectively.  \label{fig:results}}
\end{figure*}

We use the rotation curves from the Spitzer Photometry and Accurate Rotation Curves (SPARC) \citep{Lelli2016} database. The SPARC database is a sample of 175 nearby galaxies representative of the variety of galaxy types. 
We only use the 121 galaxies with high-quality rotation curves that
have 10 or more data-points. In figure \ref{fig:results} and table \ref{tab:results} we also show the results for a subsets of 106 galaxies whose inclinations are greater than $40^{\circ}$ and a subset of 71 galaxies whose inclinations are between $40^{\circ}$ and $75^{\circ}$\footnote{Edge-on and face-on galaxies labeled by SPARC as high-quality are used in the main dataset, because SPARC excludes from this category galaxies whose circular velocity is not well-measured.}. 
Since removing the more face-on and edge-one galaxies does not change the conclusions within the errors, we use the full dataset of 121 galaxies for the rest of the discussion.  Additional plots comparing various results using different subsets of galaxies based on inclination and numbers of points on rotation curves, and showing the distribution of galaxies by sampling number, can be found in the Appendix.

The SPARC database gives the measured circular velocity of the galaxies as a function of radius, $v_{obs}(r)$. The visible mass components of the galaxies are a gas disk, a stellar disk and a stellar bulge, which are also measured. As the total gravitational potential is the sum of the contributions of each mass component, it is customary to characterize the contribution of each component by a $v_i^2$ such that the sum of all $i$ components satisfies:
\begin{equation}
\label{eq:vobs}
v_{obs}^2 = \sum_{i} v_i^2.
\end{equation}
The $v_i$ are not the velocity of the mass components. They represent the contribution of the mass components to the gravitational potential and hence to the total observed velocity $v_{obs}$, via \eqref{eq:vobs}.

The SPARC database lists the following quantities as a function of the distance to the center of the galaxies:
\begin{itemize}
\item $v_{\textup{obs}}$: the observed circular velocity
\item $\sigma_{\textup{vobs}}$: the estimated uncertainty on the observed circular velocity
\item $v_\textup{gas}$: the contribution of the gas disk; $v_\textup{gas}$ is derived from the measured $HI$ gas surface densities scaled by a 1.33 factor in order to take into account the  presence of helium.
\item $v_\textup{disk}$: the contribution of the stellar disk, assuming a stellar mass-to-light ratio of 1$M_\odot/L_\odot$
\item $v_\textup{bulge}$: the contribution of the stellar bulge to the total velocity, assuming a stellar mass-to-light ratio of 1$M_\odot/L_\odot$
\end{itemize}
In converting from the observed distribution of gas and stars to their corresponding $v_i^2$ 
quoted in the SPARC database, a thin-disk approximation was used \citep{Lelli2016}.   

\begin{deluxetable*}{llll}[t]
\tablecaption{Summary of the models and their free parameters; for details see Appendix \ref{app:models}. \label{tab:models}}
\tablewidth{0pt}
\tablehead{
\colhead{Model} & \colhead{Global parameter} & \colhead{Galaxy dependent free parameter} & \colhead{Components}
}
\startdata
  CDM NFW halo & &$\Upsilon_*$, $R_s$, $\rho_0$ &Baryons, DM halo\\
  CDM pIso halo & &$\Upsilon_*$, $R_c$, $\rho_0$ &Baryons, DM halo\\
  Einasto & &$\Upsilon_*$, $R_s$, $\rho_0$, $\alpha$& Baryons, DM halo\\
  SIDM & &$\Upsilon_*$, $\sigma_{v0}$, $\rho_0$& Baryons, DM halo\\
  HIDM gas-scaling & &$\Upsilon_*$, $R_c$, $\rho_0$, $\theta$ &Baryons, DM halo, DM disk\\
  HIDM interactions-scaling & &$\Upsilon_*$, $R_c$, $\rho_0$, $\zeta$ &Baryons, DM halo, DM disk\\
  Total-baryons-scaling & &$\Upsilon_*$, $\theta_b$& Baryons, DM disk\\
  MOND & ~~~~~~~~~$a_0$&$\Upsilon_*$  &Baryons\\
  RAR  & ~~~~~~~~~$a_0$& $\Upsilon_*$ & Baryons\\
\enddata
\tablecomments{The index $i$ on the galaxy-dependent free parameter is dropped for clarity}
\end{deluxetable*}

The models we have considered are detailed in Appendix \ref{app:models}.  Each model can be written as
\begin{equation}
v^2_{\textup{model}, i} = f(v_{*i}^2, v_\textup{gas\ i}^2, r_i, \{\textup{params}_i\}, \{\textup{params}_\textup{model}\})
\end{equation}
where
\begin{equation}
v_{*i}^2=\Upsilon_\textup{disk,i} v_\textup{disk,i}^2+\Upsilon_\textup{bulge,i} v_\textup{bulge,i}^2,
\end{equation}
and $f$ is the model function, $i$ is the index of the galaxy, $\{\textup{params}_i\}$ is a set of free parameters which depend on the galaxy and $\{\textup{params}_\textup{model}\}$ are the model's free parameters.
In all the models, each galaxy is allowed to have it's own stellar mass-to-light ratio parameter $\overline\Upsilon_{*i}$ which sets the disk stellar mass-to-light ratios: $\overline\Upsilon_\textup{disk,i}=\overline\Upsilon_{*i}$.  In our baseline analysis
we take $\overline\Upsilon_\textup{bulge,i}=1.4\,\overline\Upsilon_{*i}$ as suggested by stellar population synthesis models \citep{Schombert2014}.  
The stellar mass-to-light ratios, $\overline\Upsilon_{*i}$, are constrained free parameters of the models.  We treat them the same way in each model. We verified that allowing $\Upsilon_\textup{bulge,i}$ to be a constrained free parameter does not influence the conclusions (Appendix B); this is not surprising since very few  galaxies have a significant bulge. 

Table \ref{tab:models} summarizes the free parameters of the various models.

\section{Rotation curve fitting and mass-to-light ratios}
\label{sec:ml}
The fits are done minimizing the reduced $\chi^2$ of a \{model, galaxy\} pair defined as:
\begin{align}
\chi^2_{\textup{model},i} =& \frac{1}{n_i-\nu_{mod}}\Bigg[\Big(\frac{\Upsilon_{*, i}-\overline\Upsilon_*}{\sigma_{\Upsilon_*}}\Big)^2\nonumber  \\  &+\sum_{j=1}^{n_i} \Big(\frac{v_\textup{obs, i}(r_j)-v_\textup{model, i}(r_j)}{\sigma_{v_\textup{obs, i}(r_j)}}\Big)^2\Bigg].
\label{eq:chi2}
\end{align}
Here $i$ labels the galaxy, $j$ the data point of the rotation curve; $n_i$ is the number of data points for the given galaxy and $\nu_{mod}$ is the number of degrees of freedom per galaxy of the model.   We allow $\overline\Upsilon_{*,i}$ to vary from galaxy to galaxy but deviations from the assumed mean value $\overline{\Upsilon_{*}}$ are penalized by the first term in (\ref{eq:chi2}).  
We only consider galaxies with $n_i \geq 10$, and the median and median value are 19 and 24, respectively.   
The observational measurement uncertainties used are the uncertainties assigned by SPARC for each data point.
\citep{Lelli2016, Schombert2014}.

\section{Results of Rotation Curve Fits}
\label{sec:results}

Figure \ref{fig:results} gives an overview of the quality of fits provided by the different models to the magnitude and shape of the rotation curves as a function of radius.  The upper and lower line for each model uses the full 121-galaxy or restricted 106-galaxy dataset, respectively.  The relationship between the fits provided by different models is robust, independent of whether more-face-on galaxies are excluded or not. 

The reduced $\chi^2$ is distinctly better for the HIDM models than any of the traditional models, including the SIDM and CDM models.  Interestingly, the more physical HIDM interaction-scaling (HIDM-IS) model gives a better fit than simply scaling to the gas density.  Not only is the median $\chi^2$ better than for traditional models, but the outliers are also improved.  However the empirical Einasto function does essentially as well as the HIDM-IS model. 

Subsets of models with the same number of parameters can be directly compared between themselves, e.g., the CDM and SIDM models, or the two HIDM models and the Einasto parameterization, or the MOND and RAR models.  A comparison between models with different numbers of parameters is possible using the reduced chi-squared, where the  factor $1/(n_{{\rm dof}, i} = n_i-\nu_{mod})$ in Eq. \ref{eq:chi2} disadvantages models with more free parameters.  There are typically 15-20 data points on the rotation curves of the SPARC galaxies we are fitting, and the minimum number is 10.  This means that the differences in median $\chi^2$ come from the genuinely different radial behaviors possible in the different models.

\begin{figure}[t]
\plotone{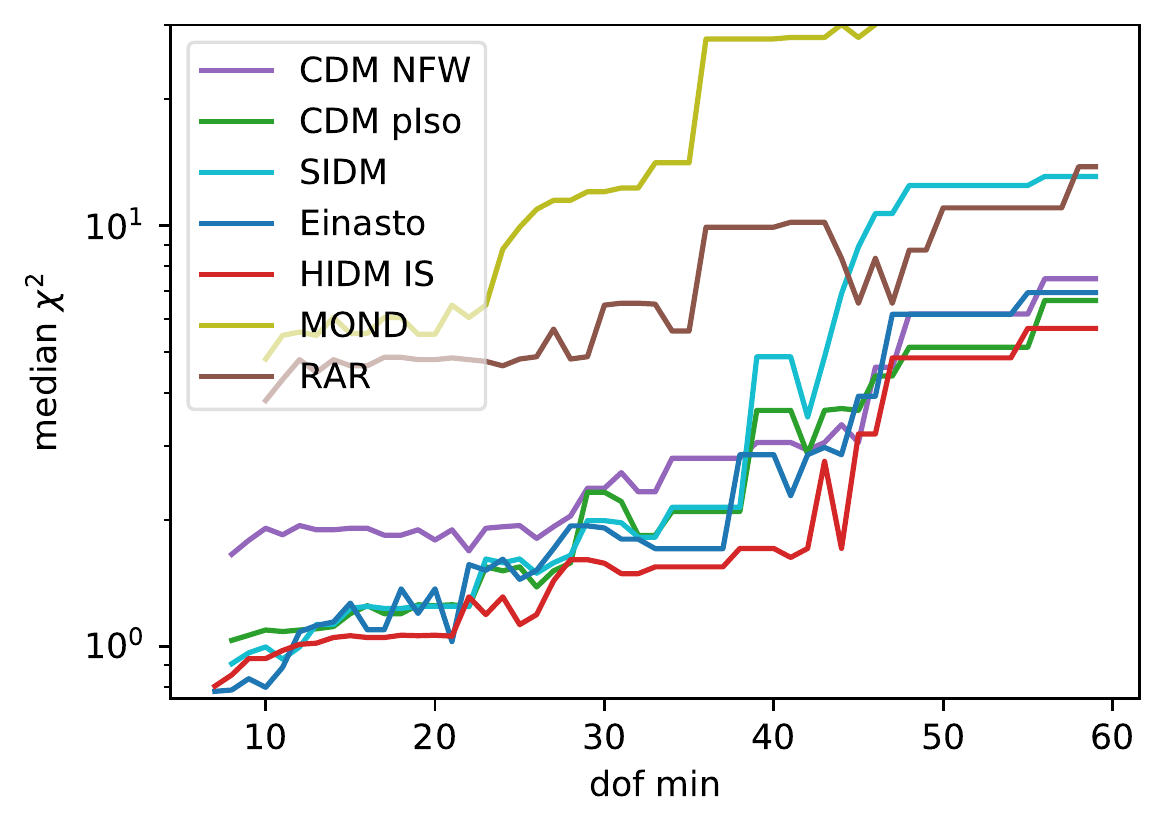}
\caption{Median reduced $\chi^2$ as a function of the minimum number of degrees of freedom in the fit: $n_{\rm dof} = n_i-\nu_{gal}$.  The relative insensitivity to the minimum number of data points required when fitting the rotation curves with the HIDM models, shows that the superiority of the HIDM fits is genuine and not an artifact of having one more parameter. It is noteworthy that Einasto is only comparably good as HIDM-IS for galaxies with small numbers of data points, and HIDM-IS is consistently best for galaxies having more precisely-sampled rotation curves. It is unsurprising that when the number of points sampled becomes large that the reduced $\chi^2$'s become systematically worse, because none of these models allow for coherent structures such as seen in the rotation curves of some highly-sampled galaxies, e.g, UGC06787 shown in the last figures in the Appendix. \label{fig:chisqvsndof}}
\end{figure}

We can directly verify that the improved fits of the HIDM relative to CDM and SIDM models is not an artifact of HIDM's having one more parameter, by examining how the median $\chi^2$ of the fits change as we increase the minimum number of data points above or below our standard criterion of at least 10.  Fig. ~\ref{fig:chisqvsndof} shows the sensitivity to min($n_{\rm dof}$) for representative models.  One sees that, apart from fluctuations, the ranking of models is preserved  independent of $n_{\rm dof}$ except that above $n_{\rm dof, min}\approx 10$ (i.e., for galaxies with more detailed rotation curves) HIDM-IS consistently out-performs Einasto and SIDM is no better than pIso. The HIDM-IS model is the overall best fitting model for essentially the entire set of galaxies, although for galaxies with fewer data-points Einasto provides an equally good or sometimes better fit.  Thus the improved $\chi^2$ of the HIDM models indicates a genuinely better description of the shape of the rotation curves than provided by traditional models. 

The significance of the differences between the median $\chi^2$'s of the different models can be quantified via a jack-knife procedure.  We divide the galaxies at random into two halves and calculate their separate median $\chi^2$'s;  repeating 1000 times, we recover the ensemble medians and standard deviation SD of the values, for each model.  Since these samples have half as many members as the full sample, we estimate a 1-sigma-like uncertainty as SD/$\sqrt{2}$; the results are shown in Table \ref{tab:results}.  
 The HIDM interactions-scaling model is 2.7-$\sigma$ better than pIso and 1.7-$\sigma$ better than SIDM, taking $\sigma$  to be the mean jacknife uncertainty of the pIso and HIDM-IS models, 0.085.  The Einasto parameterization offers a similar improvement.

\begin{deluxetable*}{lllllllll}[ht]
\tablecaption{Median reduced $\chi^2$ for the different models with jack knife uncertainty estimate. The first line corresponds to the entire dataset of galaxies with 7 or more points on the rotation cuve, the second line corresponds to the restricted dataset with inclinations between $40^\circ$ and $75^\circ$, the third line corresponds to restricting to galaxies with sufficient datapoints that $N_{dof}\geq 15$ for the given model. \label{tab:results}}
\tablewidth{0pt}
\tablehead{
\colhead{NFW} & \colhead{pIso} & \colhead{SIDM} & \colhead{Einasto} &\colhead{ HIDM-GS} &\colhead{HIDM-IS} &\colhead{TBS} &\colhead{MOND} &\colhead{RAR} 
}
\startdata
          $1.44\pm 0.17$ & $0.99\pm 0.08$ & $0.91\pm 0.08$ &
          $0.77\pm 0.07$ & $0.90\pm 0.10$ & $0.76\pm 0.09$ & $3.54\pm 0.53$ & $5.07\pm 0.71$ & $3.70\pm 0.45$ \\
          $1.63\pm 0.21$ & $1.08\pm 0.15$ & $1.12\pm 0.21$ &
          $0.78\pm 0.13$ & $1.10\pm 0.14$ & $1.04\pm 0.16$ & $3.81\pm 1.23$ & $5.40\pm 1.32$ & $5.54\pm 1.08$ \\
          $1.89\pm 0.23$ & $1.11\pm 0.16$ & $1.12\pm 0.17$ &
          $1.14\pm 0.21$ & $1.17\pm 0.17$ & $1.04\pm 0.13$ & $5.11\pm 1.14$ & $6.04\pm 1.30$ & $4.80\pm 0.66$ \\
\enddata
\end{deluxetable*}

General properties of the model fits such as the sensitivity to the mean stellar mass-to-light ratio, and examples of specific galaxies, are given in appendix \ref{app:general} and \ref{app:examples}.  Comments on the various models and their fits are given below.  

\vspace{10pt}
\subsubsection*{CDM and Einasto halos}
\begin{figure}[t]
\plotone{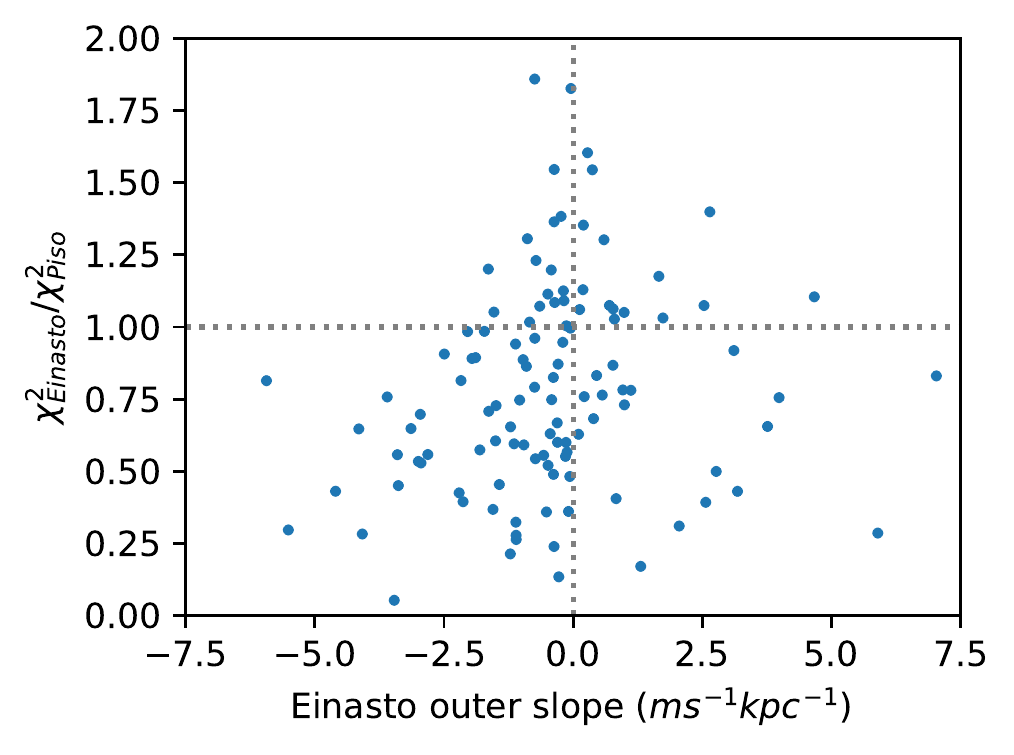}
\caption{Individual galaxies' $\chi^2_{Einasto}/\chi^2_{piso}$ vs outer slope of the Einasto halo contribution to the rotation curve, $dv_{Einasto}/dr$, evaluated at the last data point. The distribution is more populated in the lower left quadrant. These galaxies that have a dropping DM contribution at large radius are better fit by the Einasto model, while the pseudo-isothermal halo cannot model this particular feature. \label{fig:einasto_slope}}
\end{figure}

The pseudo-isothermal halo model gives a formally good fit as far as the median $\chi^2 = 0.99$ goes, with 3 free parameters per galaxy. With the same number of free parameters, the NFW halo model performs less well, with a median $\chi^2 = 1.44$ and more outliers with bad $\chi^2$. 
It has been already well known that rotation curves favor DM cores \citep{Burkert1996, Bosch2001, Gentile2004}.
The Einasto halo gives the best fits of its model category to the full dataset, with $\chi^2 = 0.77$. The more flexible functional form and additional $\alpha$ parameter allows the Einasto to reproduce some rotation curve features that cannot be modeled by the two other ``simple CDM" models. For example, Einasto fits particularly well the galaxies that have a dropping rotation curve at large radius (Fig. \ref{fig:einasto_slope}).  However it is noteworthy that the benefit of Einasto over pIso is largely restricted to galaxies with relatively few data points, as seen in Fig.~\ref{fig:chisqvsndof}. 

\subsubsection*{SIDM}
The SIDM model gives a slightly better median $\chi^2$ than the pseudo-isothermal halo model overall, but for a subset of galaxies it gives a worse fit.  It is systematically worse than pIso for galaxies with many points in their rotation curve, as seen in Fig.~\ref{fig:chisqvsndof}.  The success of the SIDM model hinges on the chosen value $\sigma_{SIDM}/m \approx 3\,{\rm cm^2/g}$, where $\sigma_{SIDM}$ is the DM self-interaction cross section at the typical relative velocity, and $m$ the DM mass.  $\sigma_{SIDM}/m$ governs the transition radius 
between the isothermal halo attributed to self-interactions and the effectively non-interacting NFW profile at large radius, with $r_1$ defined to be the radius at which there would be 1 interaction in 10 Gyr for the given $\sigma_{SIDM}$. For large enough $\sigma_{SIDM}/m$ the SIDM halos are mostly isothermal halos, which give a better fit than pseudo-isothermal.  It should be noted that in the majority of the rotation curves, the transition radius $r_1$ between the isothermal and NFW profiles is larger than the maximum radius for which the rotation curve is measured, so for these galaxies the SIDM model is equivalent to a pure isothermal halo. In some cases, matching up the NFW halos to the isothermal cores at $r_1$ produces flat outer rotation curves and this improves the fit relative to the sharper fall-off of the isothermal profile. 

\begin{figure*}
\gridline{\fig{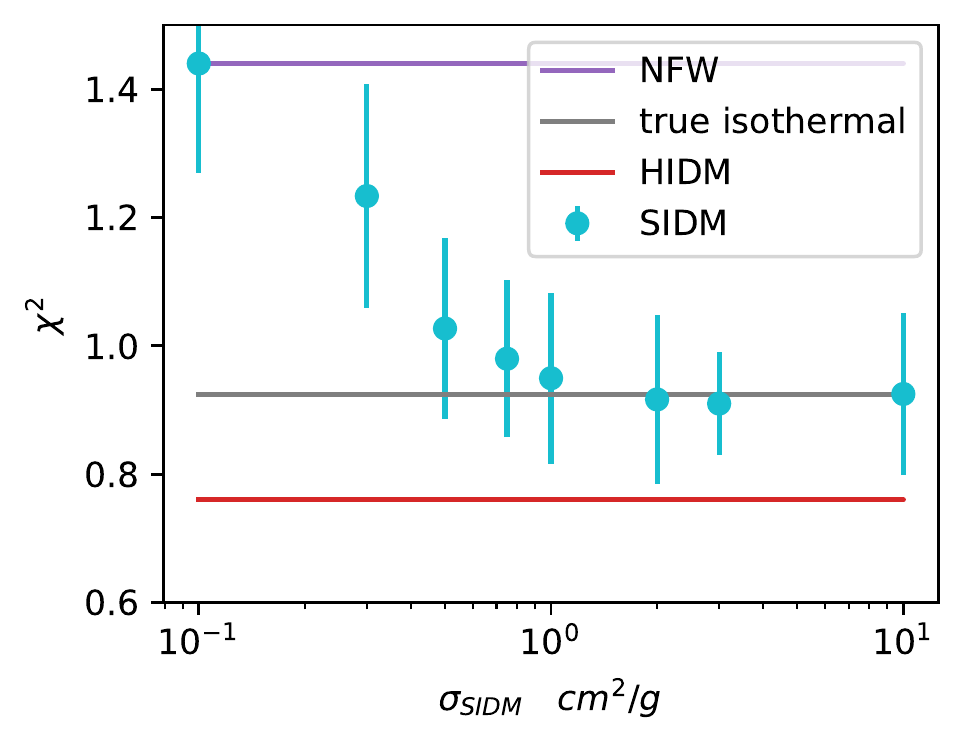}{0.4\textwidth}{}
          \fig{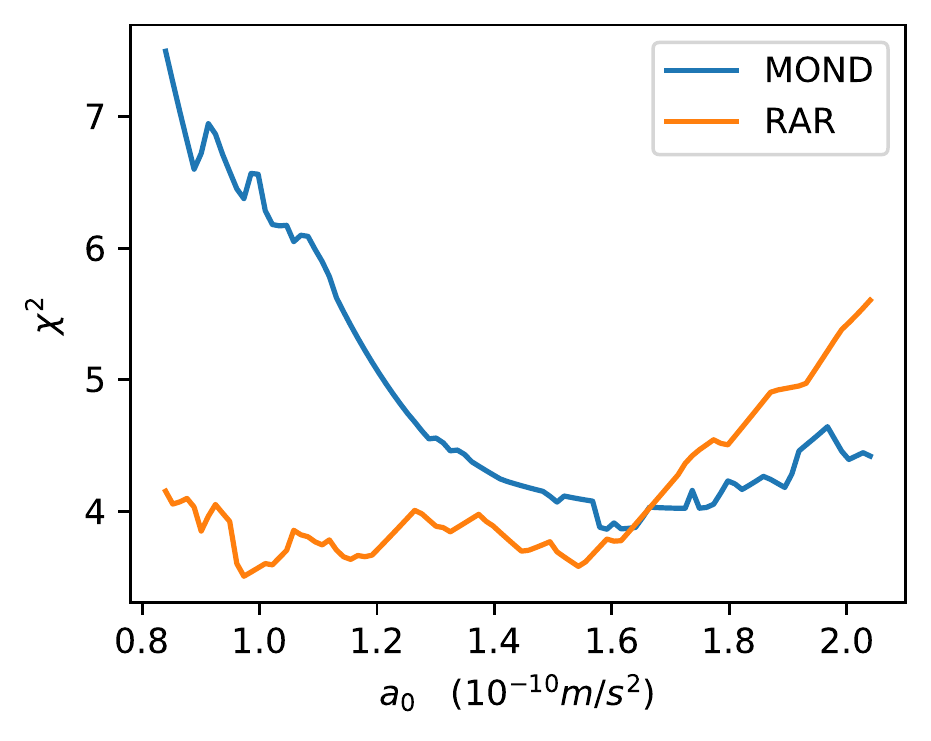}{0.4\textwidth}{}
          }
          \vspace{-0.3in}
\caption{{\it Left:}  Median $\chi^2$ of the SIDM model versus $\sigma_{SIDM}/m $. The error bars correspond to the jack knife uncertainty estimate. When $\sigma_{SIDM}/m $ is lower than $1{\rm \, cm^2/g}$, the isothermal core gets smaller and the SIDM model provides less benefit relative to NFW. 
When $\sigma_{SIDM}/m $ is too low, the SIDM fits is indistinguishable from NFW because the size of inner core is about one rotation curve data point, so the SIDM model is equivalent to a pure NFW halo. For big cross sections  ($\sigma_{SIDM}/m \gtrsim 10\, {\rm cm^2/g}$), the isothermal core is bigger than the maximum radius for which the rotation curves are measured and the SIDM model is almost equivalent to a true isothermal halo. Note that there is no noticeable improvement of the SIDM model relative to a pure isothermal halo. {\it Right:} Median $\chi^2$ of the MOND (blue) and RAR (orange) models versus acceleration scale $a_0$. We adopted $a_0=1.2\cdot 10^{-10}{\rm\, m/s}^2$ as our baseline value; note that this value is acceptable with both variants of MOND but classic MOND (Eq. \ref{eq:mond}) favors $a_0=1.6\cdot 10^{-10}\, {\rm m/s}^2$.
\label{fig:sidmchi2mond_a0}}
\end{figure*}

We went beyond the analysis reported in \citet{Ren2019}, to see if a different value of $\sigma_{SIDM}/m$ can give a better fit.   The result of our study is shown in the left panel of Fig. \ref{fig:sidmchi2mond_a0}, from which one sees that the optimal SIDM fit is achieved for $3\,{\rm cm^2/g}$, the value chosen by \citet{Ren2019}.  The fit becomes progressively worse for smaller $\sigma_{SIDM}/m$ and below $ 1 \,{\rm cm^2/g}$ is worse than pure isothermal.  For higher cross sections, the fit quality remains roughly constant with increasing $\sigma_{SIDM}/m $. It should be noted that $\sigma_{SIDM}/m  >1 \,{\rm cm^2/g}$ may be in tension with limits from the Bullet Cluster \citep{Markevitch2004}.

\subsubsection*{HIDM}
The HIDM-interaction scaling model consists of adding a DM disk scaled to the DM-gas interaction density profile.  This model gives the overall best rotation curve fits of all of the models in our study. For comparison, we also considered the HIDM-gas scaling model, in which the DM disk is simply a rescaled gas disk with the same number of free parameters as HIDM-IS.  The HIDM-GS model does not give as good a fit as the physically-motivated HIDM-IS model, but is better than DM-disk-less models (except for Einasto).  Thus the hypothesis that a DM disk forms via interaction with the gas may be valid, at least for a substantial fraction of galaxies.  We stress that our current analysis does not constrain the disk thickness; it approximates the contribution to $v^2$ as following that of the gas, which is modeled in thin-disk approximation. 

We also fitted the rotation curves with an interactions-scaling HIDM model but using a SIDM rather than pIso halo.  The median $\chi^2$ does not improve significantly (although the tail with larger-$\chi^2$ is reduced); this is compatible with the expectations of the sexaquark DM model, in which the DM self-interactions have been constrained to be too weak to have a significant astrophysical impact~\citep{Farrar2020}.

\subsubsection*{MOND, RAR and total-baryons-scaling}
To enable maximal performance for the MOND and RAR models, we allowed their parameter $a_0$ to vary;  the median $\chi^2$ of the corresponding best fits to the data are shown in the right panel of Fig. \ref{fig:sidmchi2mond_a0}. The mean $\chi^2$'s of the MOND and RAR models intersect at $a_0=1.6\cdot 10^{-10}m/s^2$, around the commonly used acceleration scale $a_0=1.2\cdot 10^{-10}m/s^2$ \citep{Scarpa2006}.  For this value they give similar quality fits, as shown in Fig. \ref{fig:sidmchi2mond_a0}. The RAR model is still acceptable for lower $a_0$ while the original MOND model is still acceptable with $a_0$ as high as $2\cdot 10^{-10}m/s^2$ (Fig.~\ref{fig:sidmchi2mond_a0}). The MOND models fail to explain the inner part of the rotation curves of the galaxies with a high central stellar density such as UGC06787 and F571-8.
The total-baryons-scaling model yields a better result than MOND, with an additional free parameter per galaxy, however it is definitely a less effective description than the CDM, SIDM and HIDM models with dark matter halos, since its median $\chi^2$ is 3-4 times higher.

\section{Interpreting the preference for HIDM models}
\label{sec:HIDMinterp}
The most striking result of the analysis presented above is the improvement in the fits to SPARC rotation curves, when the DM is not just in a traditional spherically symmetric halo, but has a disk component that is scaled to the DM-gas interactions. This does not necessarily mean that DM in galaxies has a disky component, as evidenced by the relatively successful Einasto fits\footnote{Note that the best quality Einasto fits are disproportionately for galaxies with fewer measurements in their rotation curves (Figs.~\ref{fig:chisqvsndof} and ~\ref{fig:chisqvsndof_restricted}), moreover it has not been established whether the Einasto function gives a good fit to LCDM halos in high resolution simulations, so it is not clear what significance to attach to this.}.  Even if DM does have a disky component, that would not prove that DM has non-gravitational interactions with baryons since the DM might accrete asymmetrically, on average favoring the same angular momentum axis as the baryonic disk, or the DM might ``relax" to have a disky component aligned with the baryonic disk through higher-order gravitational interactions \footnote{ Another possibility proposed by \citet{Hayashi2007} based on DM-only simulations, where the inner portion of DM halos often have prolate equipotentital surfaces oriented so the long direction is in the plane transverse to the angular momentum (where a baryonic disk would form), is that this non-trivial DM geometry could cause rotation curves interpreted with a spherical halo to {\it appear} to have a core, while actually having a (triaxial) NFW profile.}. 

In order to assess whether the improved rotation curve fits provided the HIDM-IS model could actually be due to DM-baryon interactions, we must investigate whether the values of the HIDM interactions-scaling parameter, $\{ \zeta_i \}$, are compatible with the bounds on DM-baryon interactions provided by direct detection and other limits.  This is the aim of this section.

We can contemplate two extreme possibilities for how robust a DM disk is:

{\it 1)}  The DM disk is rather fragile and is destroyed in any merger with, say, more than a 1:10 mass ratio. 

{\it 2)}  The DM disk is generally quite robust and is only significantly disturbed in a small fraction of major mergers, e.g., when the angular-momentum vectors are highly mis-aligned.

We can also consider two general scenarios for formation of the DM disk:

{\it a)} The disk builds up gradually, ``in situ'', due to collisions between DM in the halo and gas in the disk.  (In a DM-gas collision, sufficient momentum and energy are typically transferred if the DM and gas particles have comparable masses, that the post-collision DM phase-space distribution naturally approaches that of the gas \citep{Farrar2017ysn}.) In this scenario, the DM available to build up the DM disk is just the portion of the halo DM that overlaps with the gas disk.  This is the basis for the HIDM interactions-scaling parameterization as discussed in Sec.\ref{sec:HIDMis}.

{\it b)} DM-gas interactions contributing to the formation of a DM disk occur not only continuously as in {\it a)}, but also during passages of individual dwarf galaxies through the galactic plane, as they are gradually stripped of some of their stars, gas and DM; this process is documented by observed stellar streams such as the Sagitarrius stream, PAL-5 and GD-1. To first approximation, when averaged over time, this mechanism would just enhance the $\zeta$ value relative to {\it a)}.

It can happen that in some galaxies {\it a)} is dominant in determining $\zeta_i$, while in other galaxies {\it b)} produces a significant enhancement in $\zeta_i$, due to the present-day pseudo-isothermal halo  underestimating the time-averaged DM density in the gas-disk region.

\begin{figure*}
\gridline{\fig{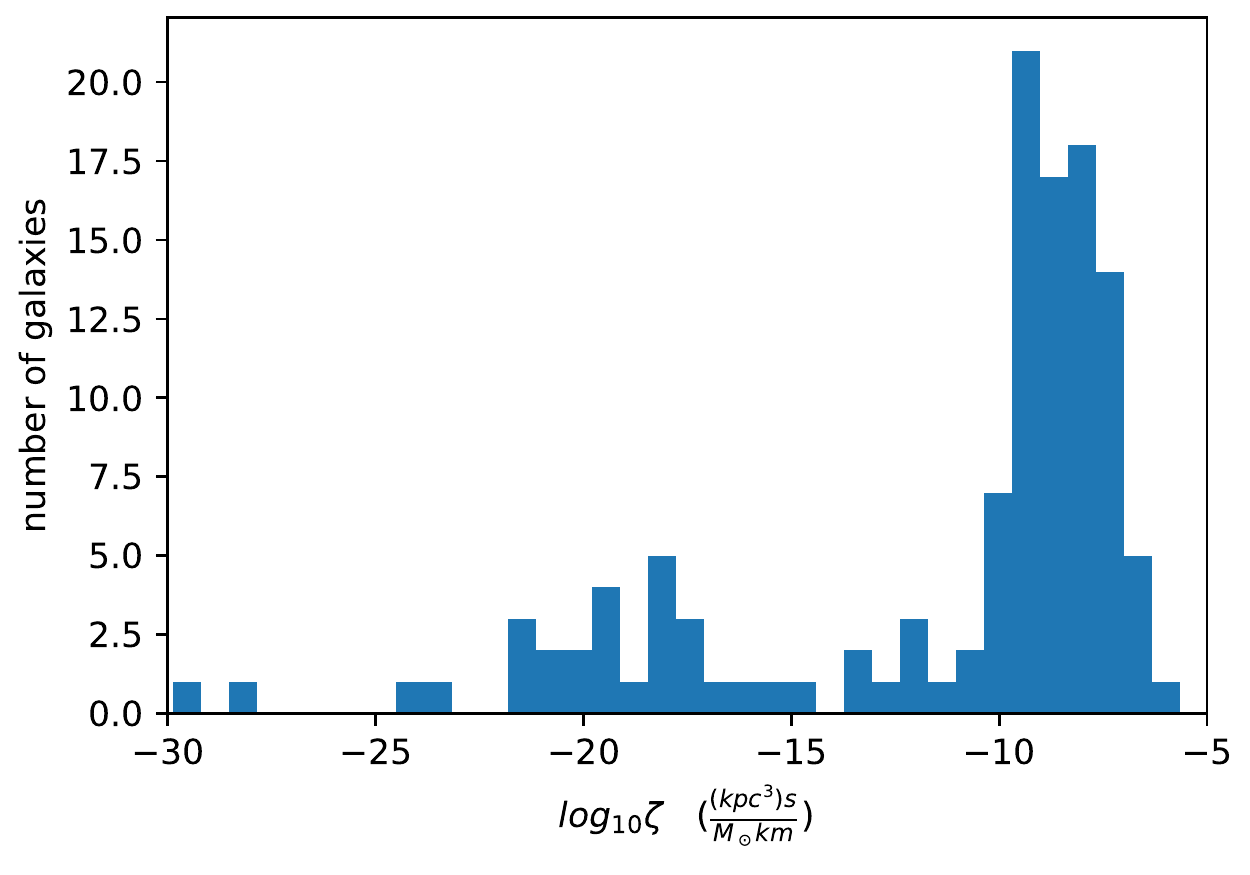}{0.4\textwidth}{}
          \fig{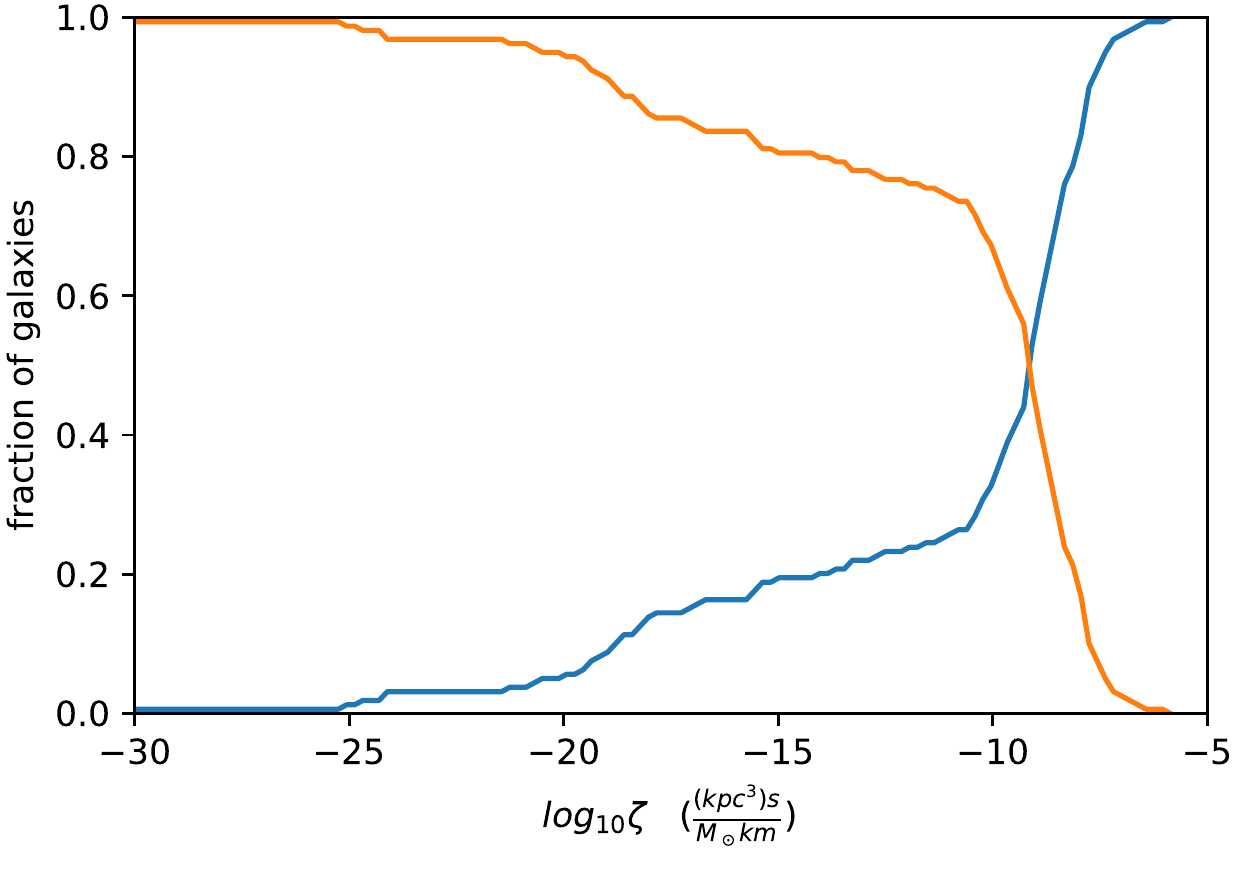}{0.4\textwidth}{}
          }
\gridline{\fig{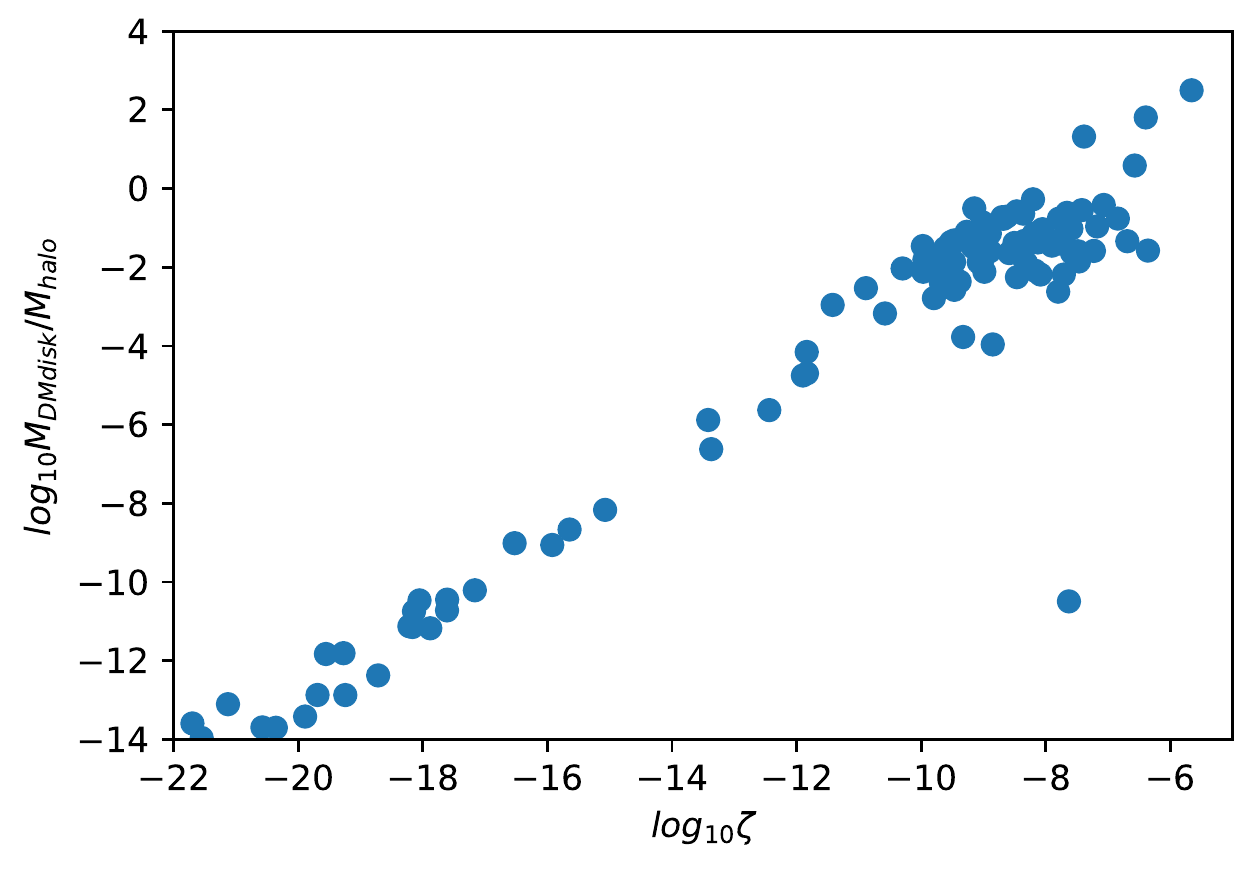}{0.4\textwidth}{}
          \fig{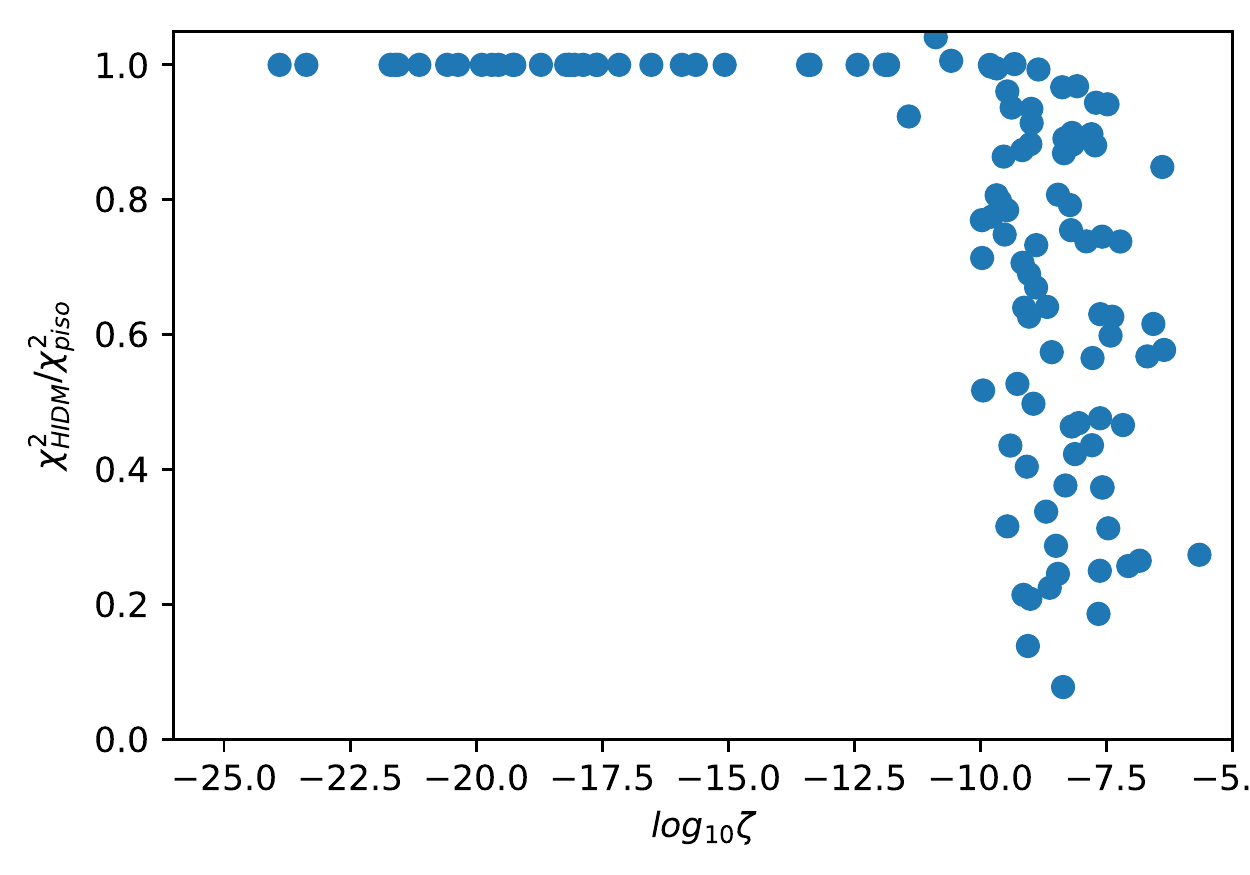}{0.4\textwidth}{}
          }  
                    \vspace{-0.3in}
\caption{{\it Upper left:} Distribution of individual galaxies' $ \zeta_i $ values.   {\it Upper right:} Cumulative distribution of individual galaxies' $ \zeta_i $ values (blue) and its complement (orange). 
 {\it Lower left:} Ratio of DM disk to DM halo mass, versus the best-fit ${\rm log}_{10}\zeta_i$.
 {\it Lower right:} Improvement in the $\chi^2$ 
for the HIDM-interactions scaling model relative to the pIso model, as a function of ${\rm log}_{10}\zeta_i$; $\zeta$ is given in units of ${\rm kpc^3 /(M_\odot km/s)}$. \label{fig:zeta}}
\end{figure*}

The upper left panel of Fig.~\ref{fig:zeta} displays a histogram of the 121 $ \zeta_i $ values for the HIDM interaction-scaling fit, with the cumulative distribution in the upper right panel.   
The lower left panel shows the ratio of the mass of the DM disk in the $i$th galaxy relative to the mass of its pseudo-isothermal DM halo, versus log $\zeta_i$. 
The lower right panel of Fig.~\ref{fig:zeta} shows the improvement in $\chi^2$
for the HIDM-interactions-scaling model relative to the same model with no DM disk (the pIso model).  

The majority of galaxies require $\{ \zeta_i \}$ in the range $(10^{-8}-10^{-9})\,{\rm kpc^3 /(M_\odot km/s)}$, and show significant improvement in the fit to rotation curves due to the DM disk.  The DM disks of these galaxies typically carry 1-10\% of the mass of the pseudo-isothermal halo, with a few galaxies calling for an even greater DM disk fraction (lower left, Fig.~\ref{fig:zeta}).
Four galaxies have a more massive DM disk than halo; their rotation curve fits are shown in Fig.~\ref{fig:massiveDMdisk} of the Appendix, making it evident how much better a fit is obtained with the DM disk in these examples.
A second population shows negligible improvement in $\chi^2$ from the presence of a disk component and has $\zeta_i \lsi 10^{-10}\,{\rm kpc^3 /(M_\odot km/s)}$. These galaxies can be interpreted as having had a DM disk-busting merger recently enough that they have not yet rebuilt a substantial DM disk.

Galaxies whose DM disks develop through continuous-disk-growth for some time $T_i$ are the inspiration for the HIDM-interactions scaling model, in the approximation that the current gas disk is typical of the average gas disk over the time $T_i$ and the circular velocity provides a fair estimate of the DM-gas relative velocity. A range of $\zeta_i$ values around the mean would arise due to time-variations in individual gas disks, and a population of higher-$\zeta$ values could also in principle be explained as instances in which a correspondingly higher fraction of the DM came into contact with gas in the disk, as would be the case if entire dwarf galaxies with all of their halo DM, passed through the gas disk multiple times before being fully merged into the halo.

From Eq. \ref{eq:SigmaDM}, we have the relationship between $\zeta_i$ and the DM-gas cross section and accumulation time $T_i$, in the idealized HIDM interactions-scaling model assuming continuous accumulation at the current rate:
\begin{equation}
\label{eq:zetaformula}
\zeta_i =  \frac{\overline{\sigma_{\rm DM}}}{\overline{m_{\rm gas}}}\, T_i~.
\end{equation}
In this expression, $\overline{\sigma_{\rm DM}}$ is the abundance-weighted DM-gas cross section, assumed in the present HIDM interaction-scaling analysis to be velocity-independent, and $\overline{m_{\rm gas}}$ is the mean mass of the gas particles, which for Galactic abundances is $2.1\times 10^{-24}$ g.  For a Yukawa interaction (as applicable for sexaquark DM and many BSM models), DM-nucleus cross sections are velocity independent except in regions around a resonance point in the Yukawa parameters where the cross-section $\sim v^{-2}$ down to $v\approx v_{\rm cst}$ below which it is a constant~\citep{Xu2021}.
Using ${\rm kpc^3 /(M_\odot km/s)} \approx 10^{26} \,{\rm cm^2 \, s \, g^{-1}}$ we can invert Eq. \ref{eq:zetaformula} to find
\begin{equation}
\label{eq:zetasigDMb}
   \overline{\sigma_{\rm DM}}  = 0.6 \, \zeta_{-9} \, T_{10}^{-1}\, 10^{-24} \, {\rm cm^2}~ ;
\end{equation}
where $\zeta_{-9} \equiv \bar{\zeta} \, 10^{9}\approx 1$ and $T = 10 \,T_{10}$ Gyr are the central values for the ensemble of galaxies. Since the neglected effects would increase the time-integrated flux, the true cross section needed will be less than this estimate.  

If DM-proton or DM-He scattering dominates $\overline{\sigma_{\rm DM}}$, Eq. \ref{eq:zetasigDMb} is (just) compatible with the latest analysis taking into account non-perturbative and finite-size effects \citep{Xu2021} using  the robust CMB-based analysis of ~\citet{Dvorkin:2013cea,Gluscevic:2017ywp,Xu:2018efh}.  The cosmological structure formation limits are stronger if Ly$\alpha$ and Milky Way satellite limits are valid, but these may be questionable, c.f., ~\citet{hotw17}. If valid, these stronger limits would exclude $\sigma_{p} \gtrsim 10^{-27} \, {\rm cm^2}$ for DM mass in the sexaquark range, $\approx 2$ GeV ~\citep{Xu2021}.  However even these stronger contraints allow a DM-nucleon Yukawa coupling parameter as large as $\alpha = 0.3$~\citep{Xu2021,Farrar2020}, which could allow DM resonant scattering on some heavier nucleus in the ISM to have such a large cross section as to satisfy Eq.~\eqref{eq:zetasigDMb} in spite of a modest fractional abundance.  To further test such a scenario requires simulations to determine the true integrated flux of DM on gas, taking into account dwarf galaxies being stripped and assimilated into the galaxy and accounting for the velocity-dependence of the dominant DM-nucleus cross section near resonance.

\section{Summary and Conclusions}

We tested 9 different dark matter and MOND-type models on the rotation curves of 121 galaxies in the SPARC database which have high-quality circular velocity measurements at 10 or more radii as well as high-quality measurements of the contribution of gas and stars to the rotation curve.  We took the stellar mass-to-light ratio of individual galaxies to be fit parameters, constraining the mean value and variance to agree with observations and stellar population synthesis modeling:  $\overline\Upsilon^* = 0.5 M_\odot/L_\odot$ and $\sigma_{\Upsilon_*}=0.25\, \overline\Upsilon_{*}$~\citep{Schombert2018}.  Requiring a realistic distribution of $\Upsilon^*$ values has not previously been done.

Fig.~\ref{fig:results} and Table \ref{tab:results} distill the results.
The MOND type models provide a very much worse description of the ensemble of galaxy rotation curves than the DM models, with median reduced $\chi^2$ a factor 3-4 larger.   
The Radial Acceleration Relation (RAR) model is only marginally better than the classic MOND model.
Pseudo-isothermal DM halos describe the DM mass distributions better than pure NFW halos, as already known. 
The self-interacting SIDM model does slightly better overall than the pseudo-isothermal model for $\sigma /m \gtrsim
 3\, {\rm cm^2/g}$, but is not as good as pseudo-isothermal for galaxies with many data points in their rotation curves.  For  $\sigma /m \lesssim \, {1\, \rm cm^2/g}$ the quality of fit rapidly degenerates, which may be problematic for the SIDM model given Bullet Cluster and other constraints.  
The SIDM halo is essentially equivalent to a pure NFW halo for $\sigma /m < 0.1 \, {\rm cm^2/g}$ and to a true isothermal halo for $\sigma /m > 10\, {\rm cm^2/g}$, however for no value of SIDM cross section does the SIDM model significantly outperform a true isothermal halo in terms of $\chi^2$ per degree of freedom.  This challenges the claim that rotation-curve fitting favors SIDM above conventional CDM as suggested by \citet{Ren2019}, although it remains an open question whether ``gastrophysics" alone can produce an isothermal halo and SIDM may be helpful for that.

The best-fitting model of the physically-motivated ones we investigated is the HIDM interactions-scaling model.  It postulates that DM interacts with baryons such that DM passing through the gas disk exchanges momentum with gas particles, leading to formation of a DM disk.  Our analysis is agnostic about the thickness of the DM disk thus formed but on physical grounds it is likely to be thick.  In the HIDM-IS model, the surface mass density of the DM disk is proportional to the DM-gas interaction probability -- the product of gas and halo DM densities times their relative velocity, approximated by the circular velocity, times cross section.  We also considered as a test a simpler ``HIDM gas-scaling" model in which the DM disk is proportional to the gas disk, with the rescaling factor a fit parameter.    If a DM disk is created by interactions with baryons, the HIDM interactions-scaling model would be the more realistic description and indeed it gives the better fit to the rotation curves.  The fact that the fit is better when a more detailed physical modeling of the process is done, suggests that DM-gas interactions may be responsible for a disky-component of the DM distribution.

As summarized in Table \ref{tab:results}, the HIDM interactions-scaling model gives a significant improvement of the fit compared to all previously proposed physically motivated models.  The reduced $\chi^2$ for HIDM interactions-scaling and Einasto models is 2.7-$\sigma$ better than for the SIDM and 1.7-$\sigma$ better than for the pseudo-isothermal model, although the relative quality of Einasto decreases for galaxies with more points on their rotation curves.  We estimated what DM-baryon effective cross-section would be needed to account for the HIDM-IS effect in a minimal accumulation scenario and found that even this upper bound on the needed cross section may be compatible with observational limits on DM-baryon scattering.  

In most previous studies of rotation curves, the DM distribution has been assumed to be spherical. We have shown that introduction of a (presumably thick) dark matter disk in addition to a spherical DM halo significantly improves the rotation curve fits in about 80\% of the galaxies.  An empirical Einasto fitting function gives as good an overall fit as the physically-motivated HIDM-IS model on the overall dataset consisting of galaxies with at least 7 points on their rotation curves, although Einasto does worse for galaxies with better-sampled rotation curves. 

Better-measured rotation curves for a larger sample of galaxies, and new analysis methods to observationally detect the presence of a thick dark matter disk or other DM asphericity, e.g., ~\citet{Loizeau2021}, are needed.
Simulations of galaxy formation including baryonic physics should be analyzed to determine whether DM disks may form by gravitational DM-baryon interactions alone, so that the success of the HIDM-IS model may not imply DM-baryon interactions. 
If DM disk formation does require HIDM interactions, simulations of galaxy assembly in the presence of HIDM interactions are needed to find the required DM-baryon effective cross-section.

\begin{acknowledgments}
We have benefited from discussions with and input from Stacy McGaugh, Marco Muzio, Digvijay Wadekar and Manoj Kaplinghat, and the suggestion of the anonymous referee to investigate possible inclination angle dependence of the results.  The research of GRF was supported in part by NSF-PHY-2013199.
\end{acknowledgments}

\newpage

\appendix

\section{The models}

\label{app:models}

\subsection{CDM: NFW, pIso and Einasto halos}
Here we model the mass distribution as a baryonic disk surrounded by a spherical DM halo.
We consider three halo distributions: a NFW \citep{NFW1996} halo, a pseudo-isothermal (pIso) halo \citep{Jimenez2003} and an Einasto halo.
The NFW halo is suggested by pure CDM simulations \citep{NFW1997} but the density diverges as r goes to 0 (i.e the distribution is cuspy).  
The pseudo-isothermal halo appears to be better adapted to describe observed DM cores and has been argued to emerge when feedback is included in the simulations.
The Einasto density function was introduced to describe stellar cluster profiles\citep{Einasto1965}, it has been first used to fit CDM halos in dissipationless n-body simulations by \cite{Navarro2004} and was recently shown to give a superior empirical description of the SPARC and THINGS rotation curve data \citep{Li2020, Chemin2011}.

The corresponding halo mass densities are given by:
\begin{equation}
\rho_{NFW}(r) = \frac{\rho_0}{\frac{r}{R_s}(1+\frac{r}{R_s})^2}
\end{equation}
\begin{equation}
\rho_{pIso}(r) = \frac{\rho_0}{(1+r/R_c)^2}
\end{equation}
\begin{equation}
    \rho_{Einasto}(r) = \rho_0 \exp\left(-\frac{2}{\alpha}\left[ \left(\frac{r}{r_s}\right)^\alpha-1 \right]\right),
\end{equation}
where $\rho_0, R_s, R_c, r_s$ and $\alpha$ parametrize the halos.
The mass inside a radius $r$ is given by:
\begin{equation}
M_{NFW}(r) = \int^r_0 4\pi r'^2\rho_{NFW}(r')dr' = 4\pi\rho_{0}\bigg[ln \Big(\frac{R_{s}+r}{R_{s}}\Big)-\frac{r}{R_{s}+r}\bigg]
\end{equation}
and
\begin{equation}
M_{Einasto}(r) = 4\pi\rho_0R_s^3e^{\frac{2}{\alpha}}\left(\frac{2}{\alpha}\right)^{-\frac{3}{\alpha}}\frac{1}{\alpha}\Gamma\left(\frac{3}{\alpha}, \frac{2}{\alpha}\left(\frac{r}{R_s}\right)^\alpha\right)
\end{equation}
where $\Gamma$ is the incomplete Gamma function.

The halo is by assumption a spherically symmetric  distribution so the contribution to the rotation curve is given by:
\begin{equation}
v_{NFW/Einasto}^2(r) = G\frac{M_{NFW/Einasto}(r)}{r} 
\end{equation}
For the pIso halo we use \citep{Jimenez2003}:
\begin{equation}
v^2_{pIso}=4\pi G \rho_{0} R_{c}^2\Big(1-\frac{R_{c}}{r} tan^{-1}\big(\frac{r}{R_{c}}\big)\Big).
\end{equation}
Both of these ``CDM" models are described by:
\begin{equation}v_\textup{model}^2 = v_{*}^2+v_\textup{gas}^2+v_\textup{halo}^2~,
\end{equation}
and both the NFW and the pIso model have 3 free parameters per galaxy: $\Upsilon_*$, $\rho_0$ the characteristic density of the halo and $R_s$ or $R_c$ the scale radius or core radius of the halo.

\begin{figure}
\centering
	\includegraphics[width=0.46\textwidth]{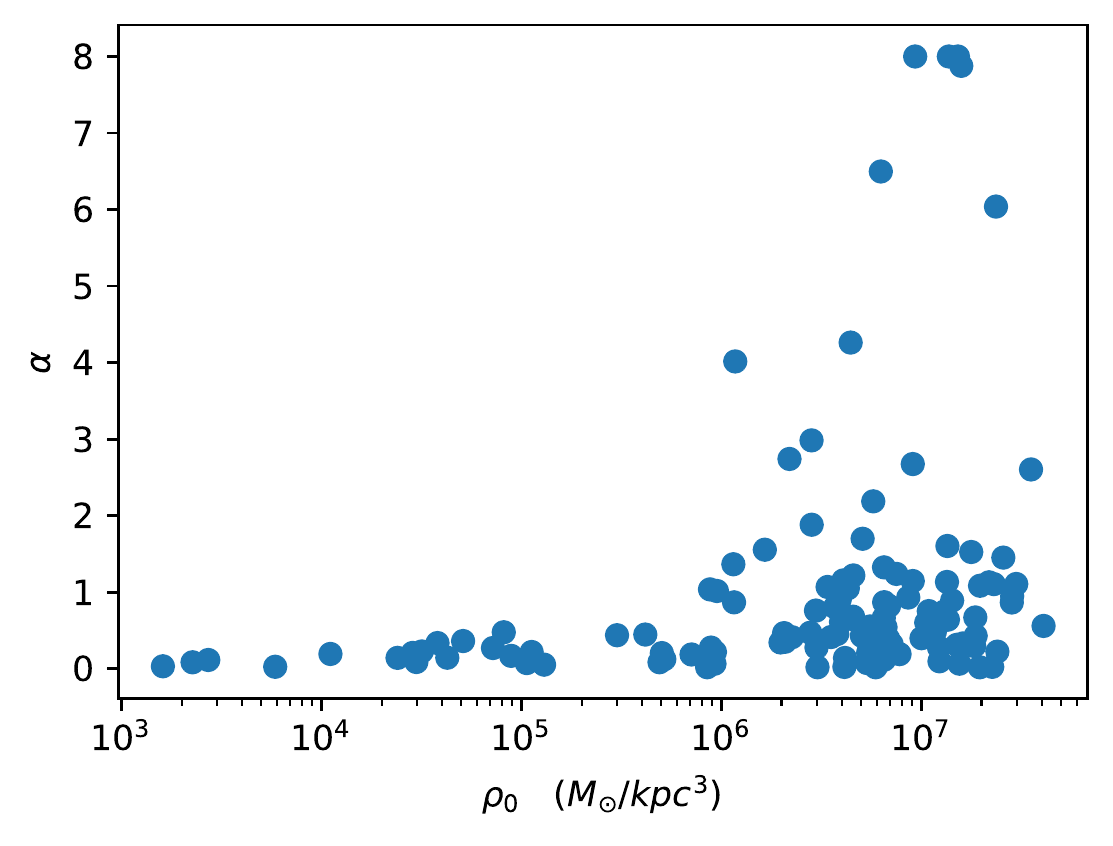}
	\includegraphics[width=0.46\textwidth]{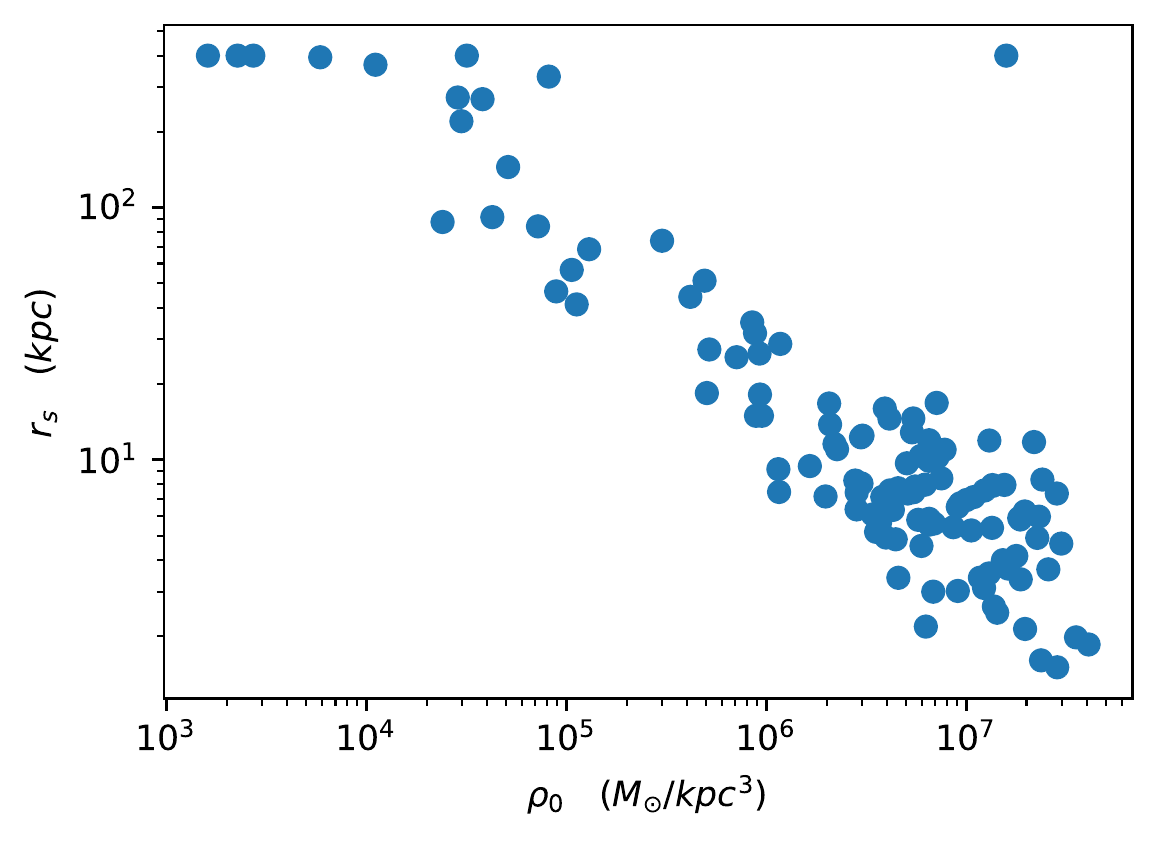}
\caption{\textit{Left}:  Scatter plot of the radial fall-off and central density parameters, $\rho_0$ and $\alpha$, best-fitting the full dataset of 121 SPARC galaxies. \textit{Right}:  Same, for the radial scale $r_0$ and $\rho_0$.   \label{fig:EinParams}}
\end{figure}

Fig.~\ref{fig:EinParams} gives additional information on the Einasto parameters found to give optimal fits to the SPARC dataset.

\subsection{Self interacting dark matter \label{sidm}}

Self-interacting Dark Matter (SIDM) has been a popular way to reconcile the smooth cores of galaxies in the face of the cuspy NFW behavior with DM only, since the seminal paper of \cite{Spergel2000}.  See also \citep{Kaplinghat2014, Kaplinghat2016}. Recently, \citet{Ren2019} have argued that SIDM provides an excellent fit for galaxy rotation curves.  In this section, we compare
the quality of such SIDM fits to the other models we consider.  

SIDM self-scattering is most prevalent in the inner part of the halo where the density is high, while it is negligible in the outer part. Hence following \citet{Ren2019}, we model SIDM halos by joining an inner isothermal halo with an outer NFW halo at $r=r_1$, interpreted as the characteristic radius of a DM particle scattered only once during the lifetime of the galaxy:
\begin{equation}
    <\sigma_{SIDM} v> \rho_{NFW}(r_1)\, t_{age}/m = 1~.
    \label{eq:r1}
\end{equation}
Here $m$ is the DM particle mass, $\sigma_{SIDM} $ is the DM self-interaction cross section and $v$ is the DM relative velocity in the halo;  $t_{age}$ is the age of the galaxy set to 10 Gyr and $\sigma_{SIDM} /m$ is set to $3\,{\rm cm^2/g}$ as in  \citet{Ren2019}.

We determine the isothermal profile by solving the Poisson equation
\begin{equation}
    \nabla^2\Phi_{tot}=4\pi G\left(\rho_{iso}+\rho_{baryons}\right)
    \label{eq:rpoisson}
\end{equation}
with $\rho_{iso}=\rho_0 e^{(\Phi(r=0)-\Phi)/\sigma_{v0}^2}$ where $\sigma_{v0}$ is the dark mater velocity dispersion.  Following \citet{Ren2019}, we treat  the baryon distribution as spherically symmetric for solving \ref{eq:rpoisson}.
The NFW halo matches to the isothermal halo at $r_1$, so that the inner mass and the densities are continuous. Hence the NFW parameters are fully determined by the isothermal halo parameter and vice versa. 
The model has 3 free parameters per galaxy: $\Upsilon_{*}$ and the isothermal halo parameters $\rho_0$ and $\sigma_{v0}$.   The self-interaction cross section $\sigma_{SIDM}$ does not appear as a parameter because as in \citet{Ren2019} we fix $\sigma_{SIDM} /m = 3\,{\rm cm^2/g}$.  We return to examine this treatment in the Results section.

\subsection{HIDM gas-scaling}
\label{sec:HIDMgs}
In the case that DM has moderate interactions with baryons, DM which passes through the disk exchanges momentum and energy with gas in the disk, resulting in a component of DM which to some extent follows the gas \citep{Farrar2017ysn}.  This motivates the model discussed in this section.  Because simulations with gastrophysics have been shown to give isothermal cores, we adopt pseudo-isothermal to be the functional form of the DM halo in both of our HIDM models \citep{Navarro1996,Chan2015}.  Thus

\begin{equation}
v_\textup{obs}^2 = v_{*}^2+v_\textup{gas}^2+v_\textup{DMdisk}^2+v_{pIso}^2
.\end{equation}
Assuming that there is a DM component which follows the gas in the disk at the level of their surface mass densities, the relation between these two is:
$\Sigma_\textup{DMdisk} = \theta \, \Sigma_{gas}$ so we have
\begin{equation}
v^2_\textup{DMdisk} = \theta \,v^2_\textup{gas}.
\end{equation}
This model has 4 free parameters per galaxy: $\Upsilon_{*}$, the gas to DM scale factor $\theta$, and the parameters of the DM isothermal halo, $\rho_0$ and $R_s$.  We emphasize that the actual DM ``disk'' represented in this analysis may be much thicker than the gas disk.

\subsection{HIDM interactions-scaling}
\label{sec:HIDMis}
This is a more physical version of the previous HIDM model, in which we account for the fact that the DM accumulates where the density of gas-DM interactions is high. If DM and gas particles have similar mass, the exchange of momentum and energy between DM and gas ejects gas into the halo and leaves DM in a more disk-like configuration~\citep{Farrar2017ysn, Wadekar2021}. DM also interacts with the gas in the hot gaseous halo, but that gas has a similar velocity dispersion as the halo DM, so such DM-gas scatterings do not modify the halo DM distribution very significantly.  The DM in the disk thus scales in proportion to the interaction rate per unit volume $\Gamma({\bf r})$ between the gas particles in the gas disk and DM particles in the DM halo: 
\begin{equation}
\label{eq:intrate}
\Gamma({\bf r}) =  n_{\rm halo}({\bf r}) \,\, n_{\rm gas}({\bf r}) \,\, \sigma_{\rm DM-gas} \,\,v_\textup{rel}({\bf r})  ~ .
\end{equation}

\begin{figure*}
\gridline{\fig{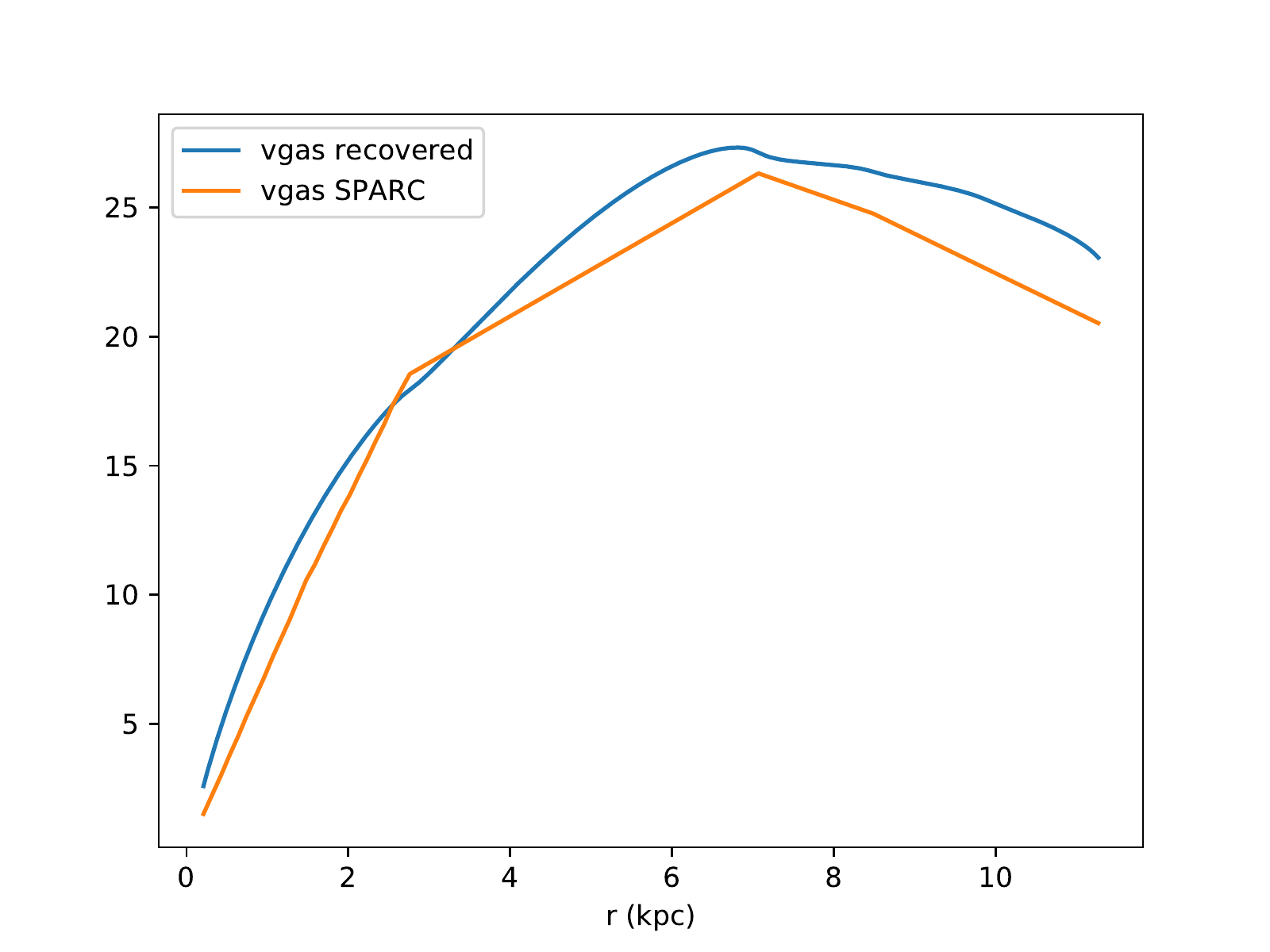}{0.3\textwidth}{(a)}
          \fig{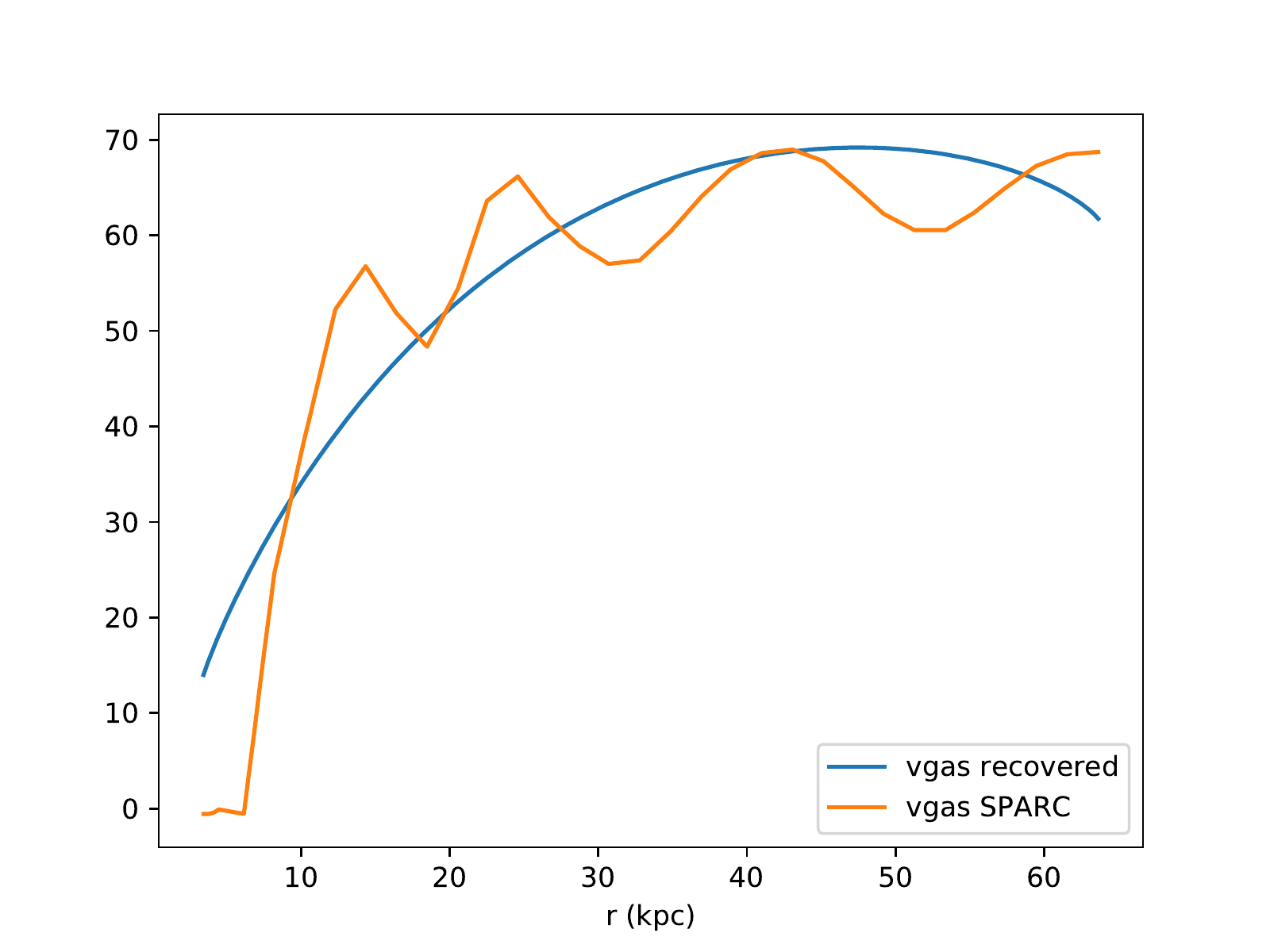}{0.3\textwidth}{(b)}
          \fig{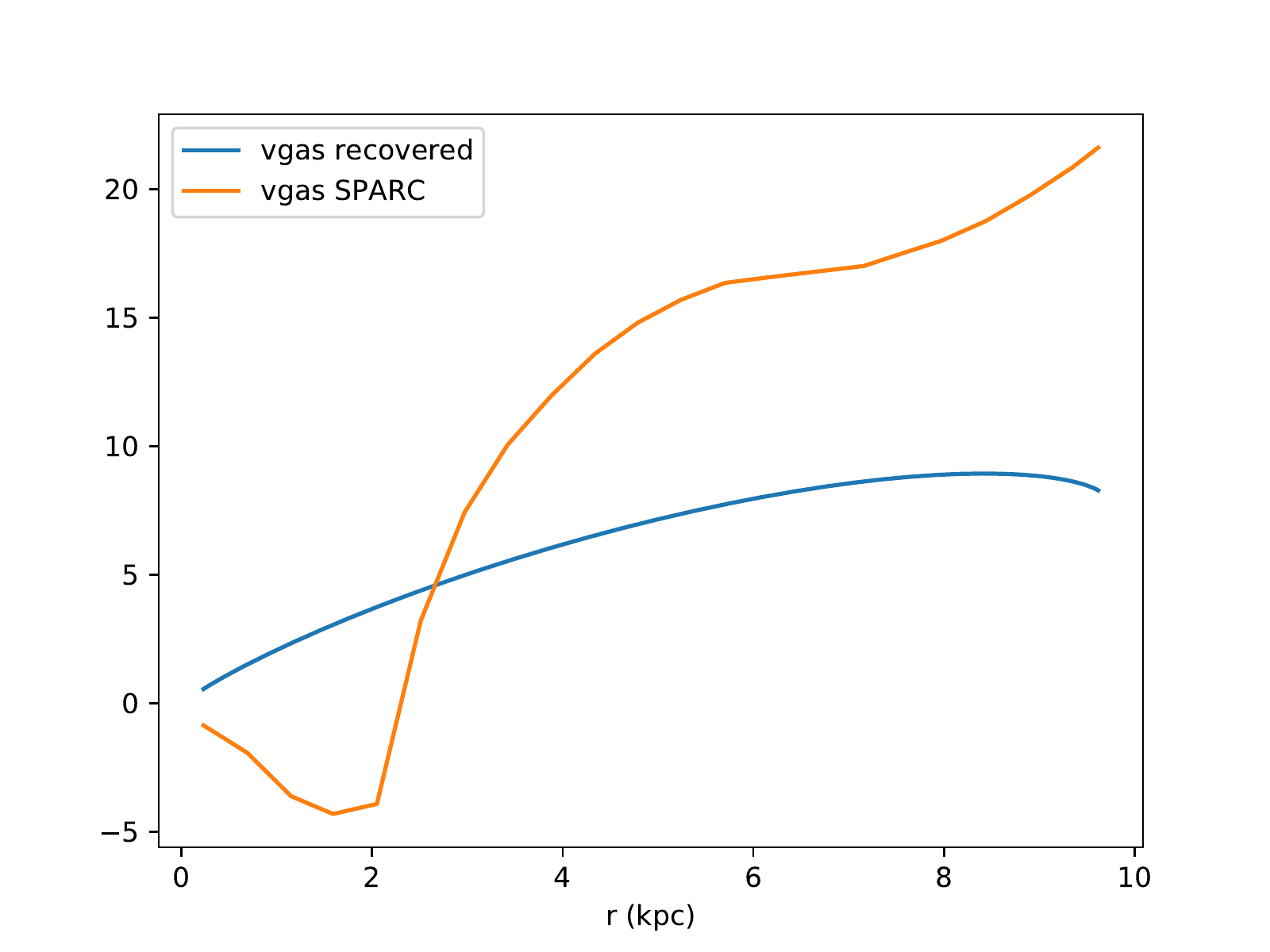}{0.3\textwidth}{(c)}
          }
\caption{\label{fig:ExpDisk} Three representative examples illustrating the fidelity of the procedure used to determine $\Sigma_\textup{gas}$ by fitting $v_\textup{gas}$ with a radially exponential disk $\Sigma_\textup{gas}(r) = \Sigma_\textup{0} e^{-r/r_0}$ \citep{Kalberla2009}, comparing input and recovered $v_\textup{gas}^2$.}
\end{figure*}

SPARC assumes a thin disk for the contribution of stellar and gas components, and we adopt the same for the scattered DM. In this approximation, the scattered DM particles produce a DM disk whose local surface number density profile is obtained by integrating \ref{eq:intrate} over $z$ and over the time $T_i$ that the DM disk has been accumulating.  
Making the simplifying assumption that the present gas distribution and DM halo density in the gas-disk region are representative of their values over the time, $T_i$, gives the DM-disk surface mass density
\begin{equation}
\label{eq:SigmaDM}
\Sigma_{\rm DM}(r)  = m_{\rm DM} \,\,  T_i \, \left( \frac{\rho_{\rm halo}(r)}{m_{\rm DM}} \right) \, \left(\frac{\Sigma_{\rm gas}(r)}{m_{\rm gas}}\right) \, \sigma_{\rm DM-gas} \,\, v_\textup{rel}(r) \equiv \zeta_i \, \rho_{\textup{halo},i}(r)\, v_{\textup{obs},i}(r)~\Sigma_{\rm gas}(r)~.
\end{equation}
In the last line we approximated $v_\textup{rel}(r)$ by the observed circular velocity at radius $r$ and took $\sigma_{\rm DM-gas}$ to be velocity independent.

Equation \ref{eq:SigmaDM} \textit{defines} the HIDM interactions-scaling model with $\zeta_i$ as the fitting parameter governing the DM disk.  In Sec. \ref{sec:HIDMinterp} we examine the relationship between the distribution of $\zeta_i$ from the fits to this model and the DM-gas cross section, but for now we treat the problem empirically with
\begin{equation}
v_\textup{model}^2 = v_{*}^2+v_\textup{gas}^2+v_\textup{DMdisk}^2+v_\textup{pIso}^2~,
\end{equation}
where, following  \citet{Lelli2016}'s treatment of the gas disk, $v_\textup{DMdisk}$ is derived from $\Sigma_\textup{DMdisk}$ by using Casertano's method, which is a way to solve Newton's equations for a mass disk \citep{Casertano1983}.

The quantities required to calculate $\Sigma_\textup{DMdisk}$ are $\rho_\textup{pIso}$, $v_\textup{obs}$ and $\Sigma_\textup{gas}$; $\rho_\textup{pIso}$ is determined by the fit, while $v_\textup{obs}$ and $\Sigma_\textup{gas}$ are observed data. However we do not have systematic access to the radial surface gas densities of the SPARC dataset. In order to compensate for this lack of underlying data, we recover $\Sigma_\textup{gas}$ by fitting $v_\textup{gas}$ with a radially exponential disk $\Sigma_\textup{gas}(r) = \Sigma_\textup{0} e^{-r/r_0}$ \citep{Kalberla2009}.  A few examples of the fidelity of this procedure are given in the Fig.~\ref{fig:ExpDisk}, where we see that the recovered $v_\textup{gas}^2$ generally agree to $\approx$20\%, although small scale structure is lost.  Hopefully the underlying $\Sigma_{\rm gas}$ data will become publicly available in the future.    

This model has 4 free parameters per galaxy: $\Upsilon_{*}$,  $\zeta$ the DM disk interaction-scaling factor, and the parameters of the halo $\rho_0$ and $R_c$.
\subsection{Total-baryons-scaling}
\cite{Swaters2012} fit a set of rotation curves with the model
\begin{equation} 
\label{eq:maxbar}
v_\textup{obs,i}^2 = \Upsilon_\textup{disk,i} \,  v_\textup{disk,i}^2 + \Upsilon_\textup{bulge,i} \, v_\textup{bulge,i}^2+\eta_i\, v_{gas,i}^2.
\end{equation}
Here $\Upsilon_\textup{disk,i}$ $\Upsilon_\textup{bulge,i}$ and $\eta_i$ are unconstrained free parameters. The resultant fits work well for the majority of the galaxies in their data set. They argue this means DM could follow the baryons in those galaxies and an extended halo is not needed (if the fit parameters $\Upsilon_\textup{disk,i}$, $\Upsilon_\textup{bulge,i}$ and  $\eta_i$ physically are reasonable). However, this fit does not work for a small subset of galaxies. A possible hypothesis to account for that is that those galaxies are recent mergers and the DM is located primarily in a halo rather than following the gas in the disk.

To quantitatively compare the Swaters et al. 2012 model with other models, we study a similar model which scales the total baryon density (bulge, stellar disk and gas) with a single overall scaling factor $\theta_b$. Effectively this assumes that there exists a DM component which exactly follows the baryons in the galaxy.  (This total-baryons-scaling model could also arise in an extreme case of hadronically interacting DM, where the DM initially in a halo would relax to follow the baryonic distribution, including baryons in stars which themselves formed from gas \citep{Farrar2017ysn}.)  The relation between the DM and stellar and gas surface mass densities is then
$\Sigma_\textup{DMdisk} = \theta_b \,(\Sigma_*+\Sigma_\textup{gas})$.    As $v^2$ is proportional to the gravitational potential, scaling the densities is equivalent to scaling $v^2$. Hence,
\begin{equation}
v_\textup{obs}^2 = v_{*}^2+v_\textup{gas}^2+v_\textup{DMdisk}^2
\end{equation}
with
\begin{equation}
v^2_\textup{DMdisk} = \theta_{b} \,(v^2_{*}+v^2_\textup{gas}).
\end{equation}
This model has 2 free parameters per galaxy: $\theta_b$ and $\Upsilon_{*}$, entering through  $v_*$.

\subsection{Modified Newtonian Dynamics (MOND) and Radial Acceleration Relation (RAR)}

The acceleration $a$ in the MOND theory is related to the Newtonian acceleration $a_N$ by a new fundamental parameter $a_0$ and an interpolation function $\mu$ \citep{Scarpa2006}:
\begin{equation}
\frac{a_N}{a} = \mu\left(\frac{a}{a_0}\right)
\end{equation}
where $\mu(x)=\frac{x}{1+x^2}$. Solving for $a$ gives
\begin{equation}
a = a_N \sqrt[]{\frac{1}{2}+\frac{1}{2}\sqrt[]{1+\Big(\frac{2a_0}{a_N}\Big)^2}}.
\label{eq:mond}
\end{equation}
Thus, the contribution to the rotation curve is

\begin{equation}
v^2_{MOND} = \frac{v^2_N}{r} \sqrt[]{\frac{1}{2}+\frac{1}{2}\sqrt[]{1+\Big(\frac{2a_0 r }{v^2_N}\Big)^2}}
\end{equation}
where
\begin{equation}
v_N^2 = v_{gas}^2+v_{*}^2.
\end{equation}
$a_0$ is usually set as $1.2\cdot 10^{-10}\textup{m\,s}^{-2}$ \citep{Scarpa2006}.

In \citep{McGaugh2016} the existence of a radial acceleration relation (RAR) is shown and the observed acceleration is related to the baryonic acceleration by:
\begin{equation}
a = a_N\left(1-e^{-\sqrt{a_N/a_0}}\right)^{-1}.\label{eq:rar}
\end{equation}
This equation can be considered as a new empirical interpolation function that well describes the data.

We fit both the classic MOND and RAR models to the SPARC rotation curves, first for the standard $a_0=1.2\cdot 10^{-10}\, \textup{m\,s}^{-2}$ and then for different $a_0$ values. For convenience, we call Eq. (\ref{eq:mond}) MOND and Eq. (\ref{eq:rar}) RAR; $\Upsilon_{*}$ is the only free parameter for each galaxy.

\section{Stellar mass-to-light ratios  }
The determination of the stellar mass-to-light ratio of a particular galaxy, $\Upsilon_{*,i}$, is a critical step in rotation curve fitting.
The standard approach in previous works has been either to leave $\Upsilon_{*,i}$ as a free parameter of the fit for each galaxy, leading to unphysical resultant distributions of $\Upsilon_{*,i}$, or to fix it based on stellar population synthesis models (e.g., \cite{Chemin2011}).
In this work, we allow $\overline\Upsilon_{*,i}$ to vary from galaxy to galaxy but we penalize deviations from the assumed mean value $\overline{\Upsilon_{*}}$ by adding a term to the $\chi^2$ in Eq. (\ref{eq:chi2}).
Upper bounds on $\Upsilon_{*}$ can be derived from the ``maximum stellar disk" fit where one tries to maximize the contribution of the stars to the rotation curve. The mean maximum-disk mass-to-light ratio for SPARC galaxies is $\overline{\Upsilon_{*}^{max}} \approx 0.7 M_\odot/L_\odot$ \citep{Lelli2016}. (Mass-to-light ratios for SPARC galaxies are quoted at $3.6\mu m$.) This is substantially higher than the estimated stellar mass-to-light ratio reported in the DiskMass Survey (DMS) which gives $\overline\Upsilon_{*} \approx 0.2 M_\odot/L_\odot$ \citep{Swaters2014}. Stellar population synthesis 
models report mean values between 0.4 and 0.6 $M_\odot/L_\odot$ \citep{Schombert2014, McGaugh2014, Meidt2014}.
 Here we assume the mean stellar mass-to-light ratio of the SPARC galaxies is $\overline\Upsilon_* = 0.5 M_\odot/L_\odot$ as suggested by \citet{Schombert2018} using data from the main sequence (stellar mass versus stellar formation rate) and stellar population models.  
We take $\sigma_{\Upsilon_*}=0.25\, \overline\Upsilon_{*}$ from \citet{Schombert2018}. For the galaxies which have a bulge, we set $\Upsilon_\textup{bulge} = 1.4 \,\Upsilon_{*}$, as suggested by stellar population synthesis models.

Each model fit returns the stellar mass-to-light ratio $\Upsilon_{*,i}$ for each galaxy; the distribution of $\Upsilon_{*,i}$ values is shown in the right panel of Fig \ref{fig:ml_mass}. 
The performance in terms of $\chi^2$ of the tested models is reflected in the distribution of mass-to-light ratios. 
DM model fits have a mass-to-light ratio distribution peaked at around 0.5, the assumed average value based on stellar population synthesis models. However MOND predicts a maximum of the $\Upsilon_*$ distribution around 0.7, significantly larger than inferred from stellar population synthesis modeling \citep{Schombert2014, McGaugh2014, Meidt2014}.

We adopted $\overline\Upsilon_*=0.5$ as the mean value for the study, but to check sensitivity of the conclusions to this chosen value, we redid the analysis for a range of mean mass-to-light ratios between $0.1$ and $1$.  How the overall quality of the rotation curve fitting depends on $\overline\Upsilon_*$ is shown in the right panel of Fig.~\ref{fig:ml_mass} for illustrative cases.  One sees that the HIDM models' quality of fit is insensitive to the assumed mean $\overline\Upsilon_*$, below $\approx 0.6$ while the CDM models have a stronger preference for some particular $\overline\Upsilon_*$. 
We also find that treating $\Upsilon_\textup{bulge}$ as a constrained free parameter with mean $1.4\overline\Upsilon_*$ and spread $0.25\overline\Upsilon_*$ does not significantly impact our results. For example, free (fixed) $\Upsilon_\textup{bulge}$ gives $\chi^2_{NFW}=1.44 (1.40)$ and $\chi^2_{pIso}=0.98 (0.99)$. Note that $24\%$ of the galaxies in the dataset have a bulge.

\citet{Li2020} does MCMC fitting allowing the galaxy distances and inclinations to vary from the tabulated SPARC values.  We decided not to do this because we do not have access to the underlying data needed for a careful analysis: a change in inclination leads to a simple rescalling of $v_{obs}$ but would have non-trivial effects on the inferred $v_{gas}$ and $v_*$.
Nor can we make use of the \cite{Li2020} values, since they were optimized under the assumption of specific DM models and we have different models.  We could consider optimizing the distance independently for each model.  Both methods are in good agreement since we recover $\chi^2_{\text{Einasto}} < \chi^2_{\text{pIso}} < \chi^2_{\text{NFW}}$ as in \cite{Li2020} where for example, $88\%$ of the pIso fits have $\mid\frac{D-D_{\text{SPARC}}}{D_{\text{SPARC}}}\mid$ and $92\%$ have $\mid\frac{\sin(i)-\sin(i_{\text{SPARC}})}{\sin(i_{\text{SPARC}})}\mid$ with $D$ the fit distance, $D_{\text{SPARC}}$ the tabulated distance, $i$ the fit inclination and $i_{\text{SPARC}}$ the tabulated inclination. These inclination and distance adjustments lead to changes in $v_{obs}$ that are small compared to the error $dv_{obs}$ given in SPARC.

\section{General properties of the model fits}
\label{app:general}

\begin{figure*}
\gridline{\fig{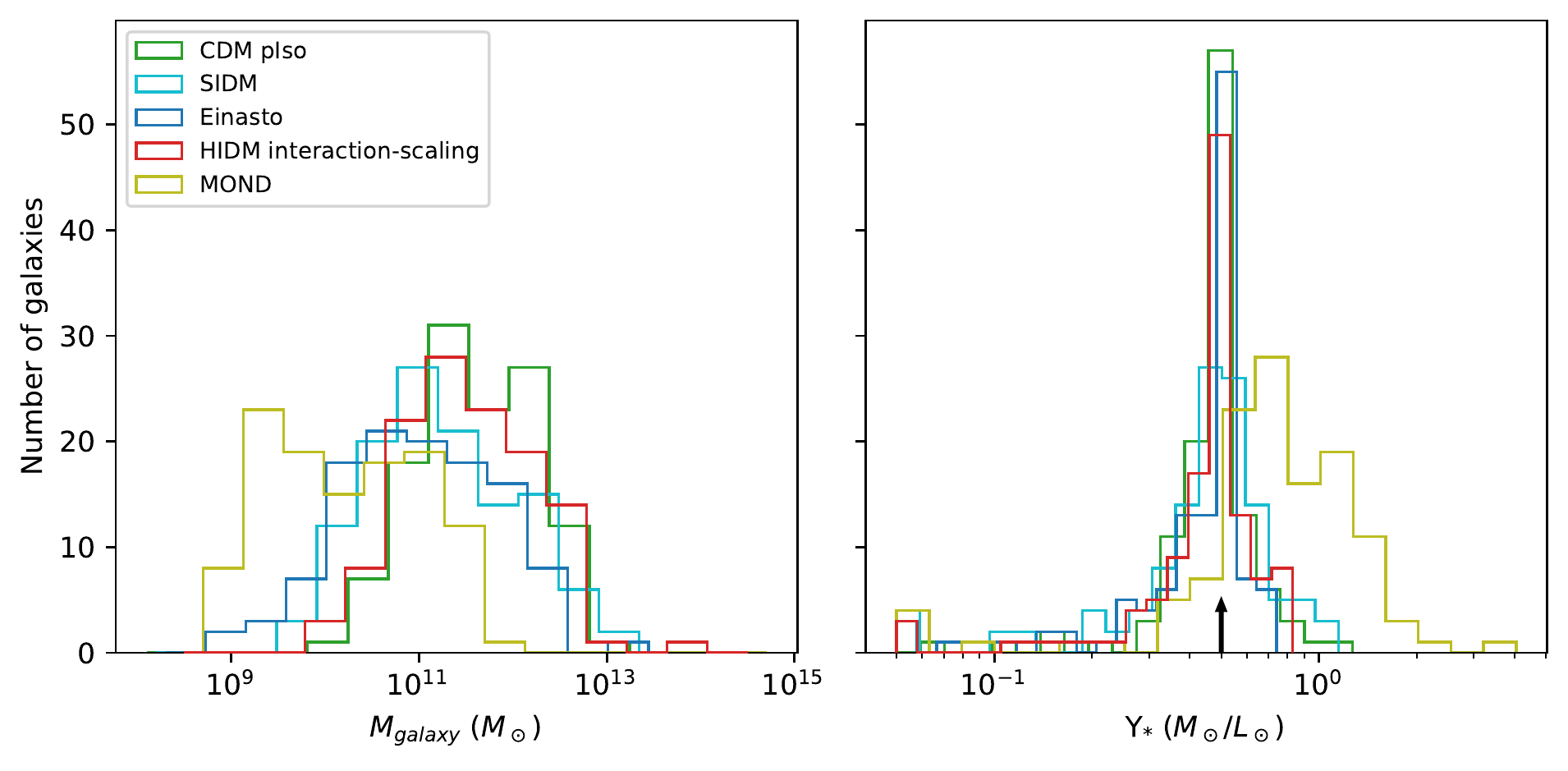}{0.55\textwidth}{}
          \fig{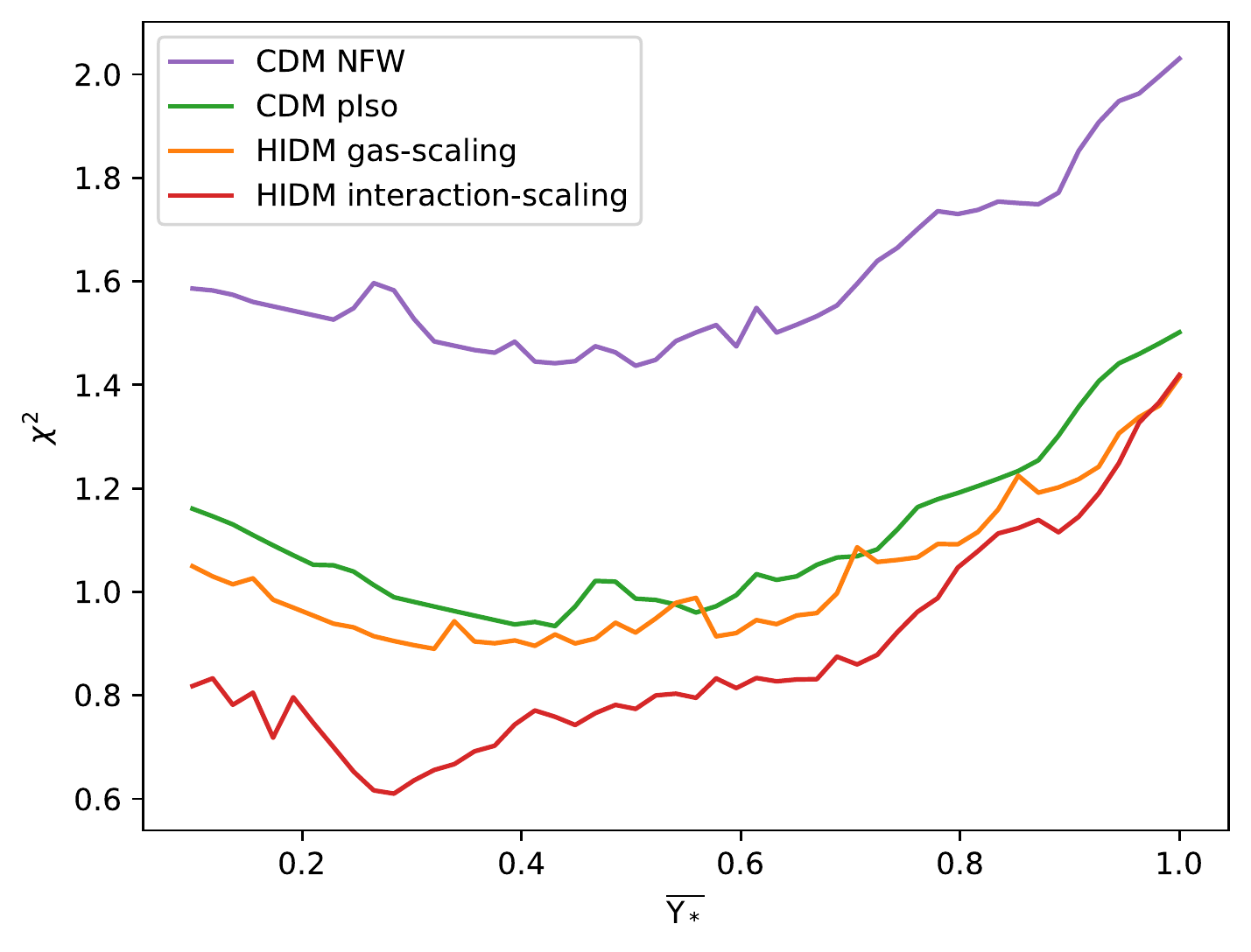}{0.35\textwidth}{}
          }
\caption{\textit{Left}: Histogram of the galaxy masses according to representative models. \textit{Center}: Histogram of the stellar mass-to-light ratios derived from the different models. The arrow represents $\Upsilon_*=0.5$, the assumed mean value. \textit{Right}: Median $\chi^2_{\rm dof}$ versus assumed mean mass-to-light ratio in the fitting.  $\overline\Upsilon_*=0.5$ is compatible with all the halo models, in agreement with \cite{Schombert2018} estimations. \label{fig:ml_mass}}
\end{figure*}

\begin{figure*}[t]
\plotone{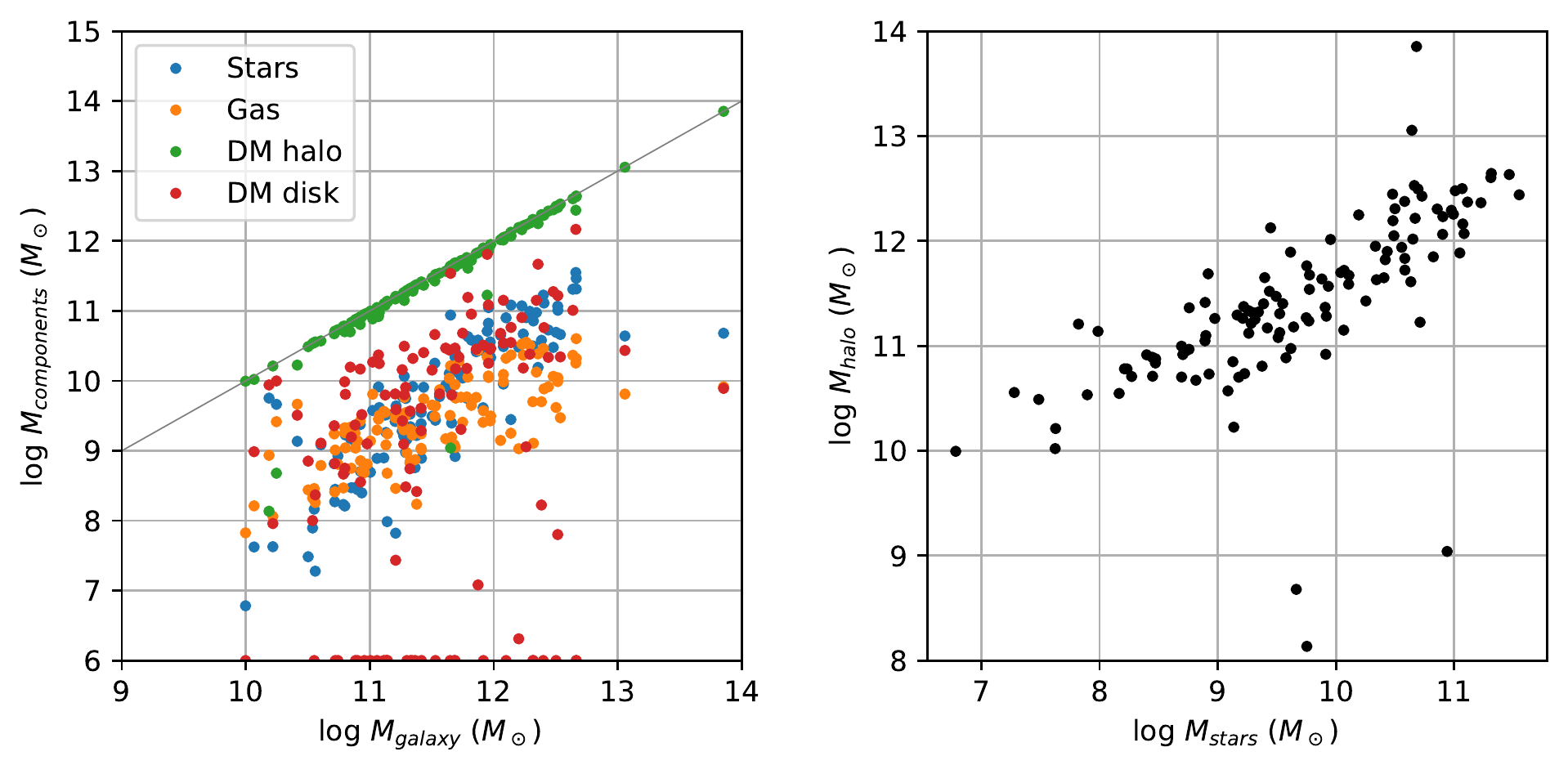}
\caption{\textit{Left}: Mass of the four components of the HIDM interactions-scaling model versus the total mass of the galaxies. One galaxy is represented by 4 points.  \textit{Right}: Stellar mass to halo mass relation for the HIDM interactions-scaling model. \label{fig:massdistrib}}
\end{figure*}

The fits enable us to estimate the total mass of each of the galaxies in the sample using the inferred amount of dark matter.  The left panel of Fig. \ref{fig:ml_mass} shows the resultant distribution of total galaxy masses for each of the models.  
Fig. \ref{fig:massdistrib} displays some important inferred physical features of galaxies, for the overall best-fitting model, HIDM-interaction-scaling.  The plots for the HIDM gas-scaling model are similar.  The left panel of Fig. \ref{fig:massdistrib} shows how the total mass of each galaxy is distributed between the stars, the gas and the two components of the dark matter.  The distribution of the masses of the DM disks is similar to the distribution of the masses of the gas disks, with a median ratio of 2.0.  The right panel of Fig. \ref{fig:massdistrib} shows the correlation between stellar and halo mass.  According to the HIDM models, the median stellar mass to halo mass ratio (SMHR) is 1.8\%.  This value is consistent with estimations from the VIMOS Ultra Deep Survey, which finds that the SMHR ranges from 1\% to 2.5\%  for redshifts $2 < z < 5$ \citep{Durkalec2014}.

\begin{figure}
\centering
	\includegraphics[width=0.32\textwidth,trim = 0.9cm 0 0 0, clip]{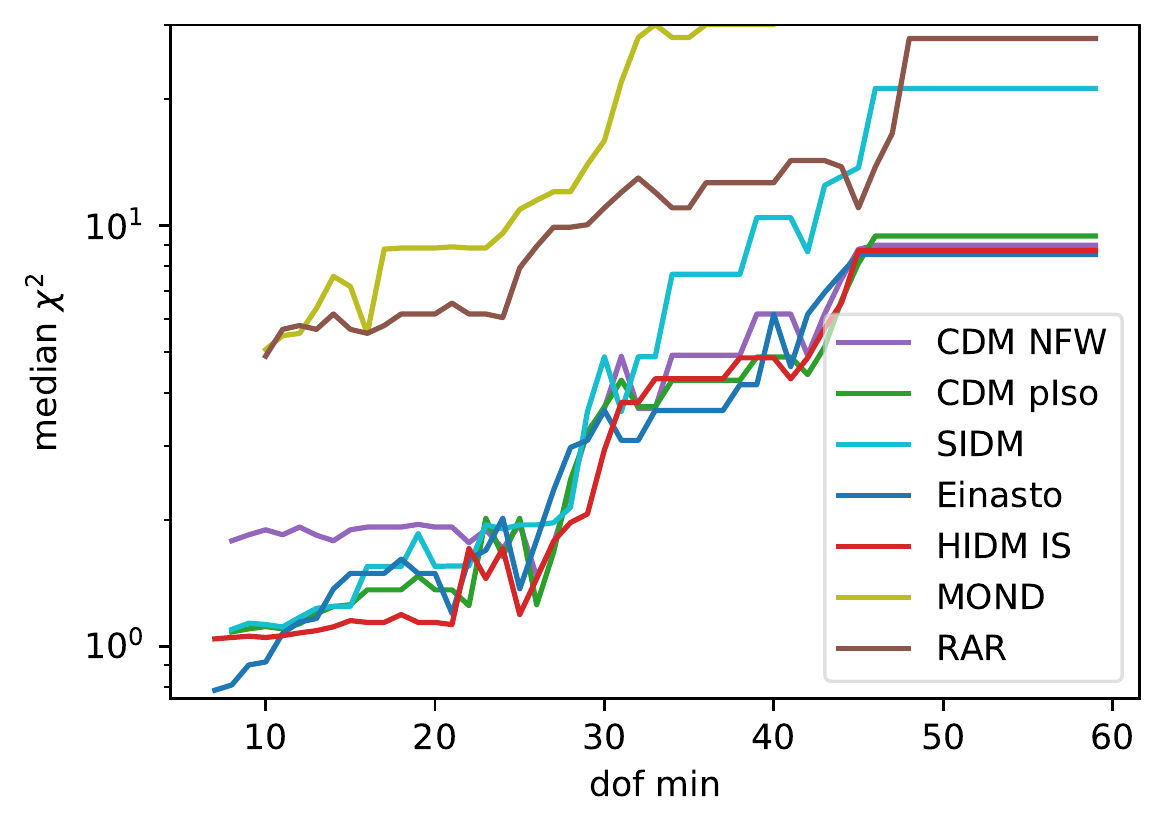}
	\includegraphics[width=0.36\textwidth]{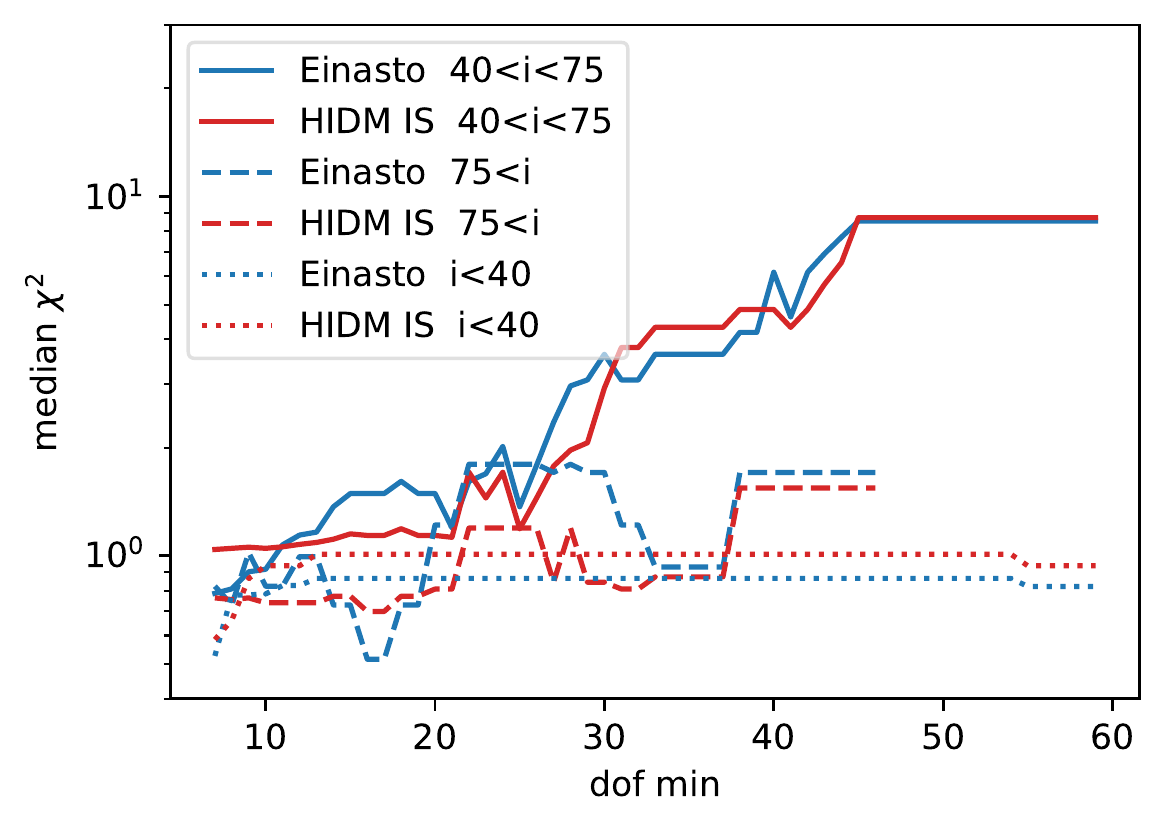}
	\includegraphics[width=0.29\textwidth]{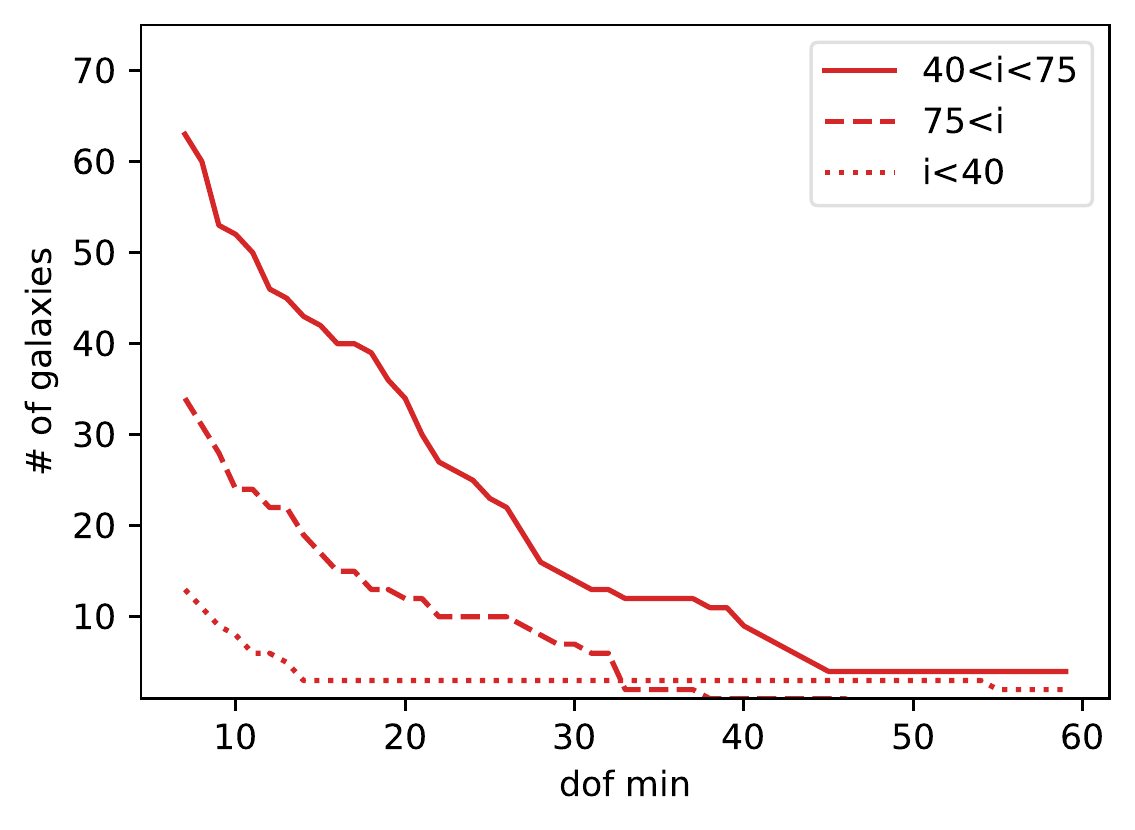}
\caption{\textit{Left}:  Median reduced $\chi^2$ as a function of the minimum number of degrees of freedom in the fit: $n_{\rm dof} = n_i-\nu_{gal}$ for the restricted dataset with inclinations between $40^\circ$ and $75^\circ$.  \textit{Center}:  Same, for the Einasto and HIDM-IS models, separated by inclination angle region. \textit{Right}:  Number of galaxies in the 3 inclination-angle-bins vs. $n_{\rm dof,min}$  \label{fig:chisqvsndof_restricted}}
\end{figure}

We also display in Fig.~\ref{fig:chisqvsndof_restricted} some facets of the inclination-angle and minimum-$n_{\rm dof}$ dependence of the results, concentrating on HIDM-IS and Einasto, the two most successful models.

\section{Specific example galaxies}
\label{app:examples}
\begin{figure}[H]
\plotone{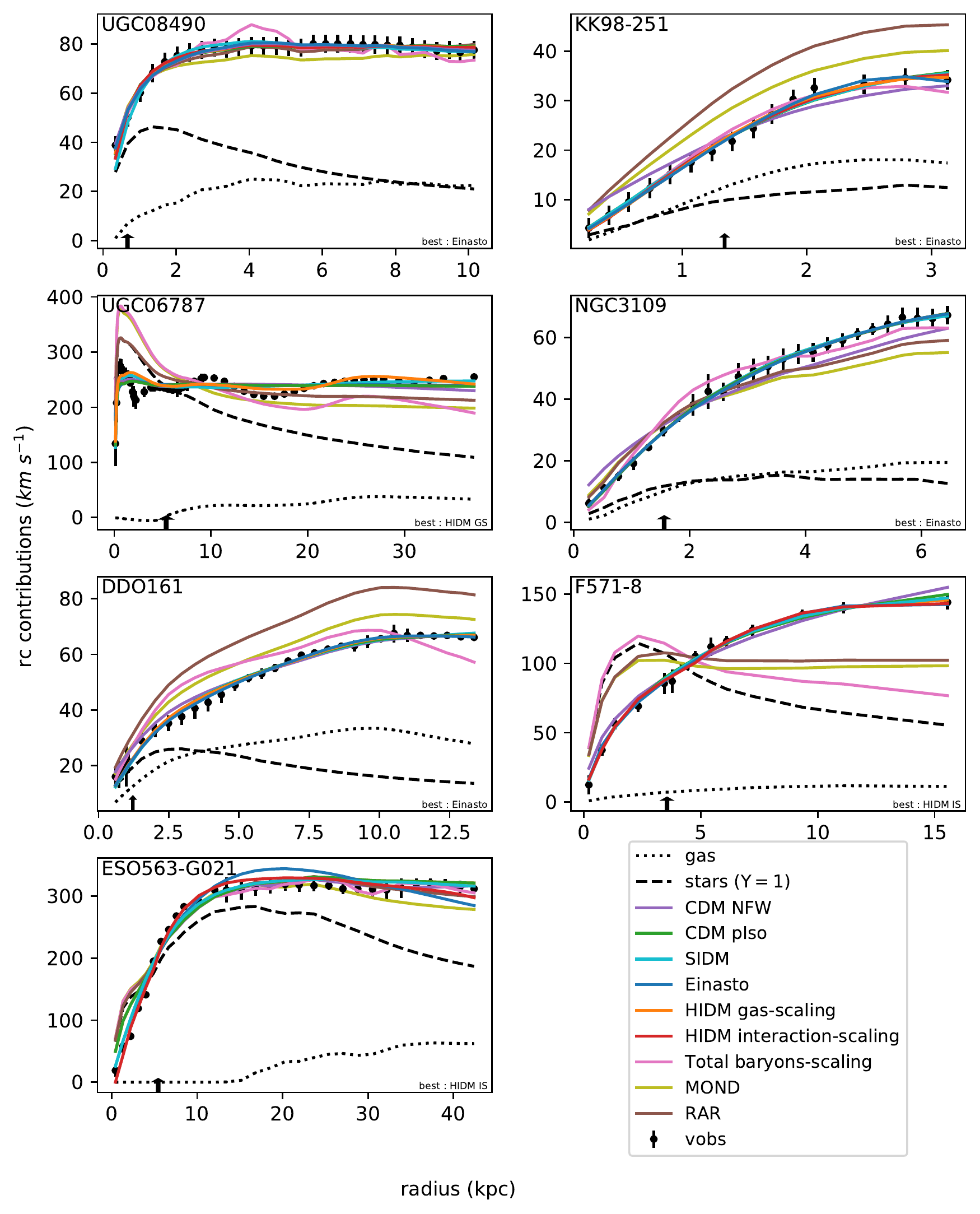}
\caption{ \label{fig:RC7}Rotation curve data for 7 illustrative galaxies. The dashed line is the contribution of the stars for $\Upsilon_*=1M_\odot/L_\odot$ and the doted line is the contribution of the gas disk. Colored lines are the best-fit rotation curves for the 8 model. The black arrow is the disk scale length \citep{Lelli2016}}
\end{figure}

\begin{figure}[H]
\plotone{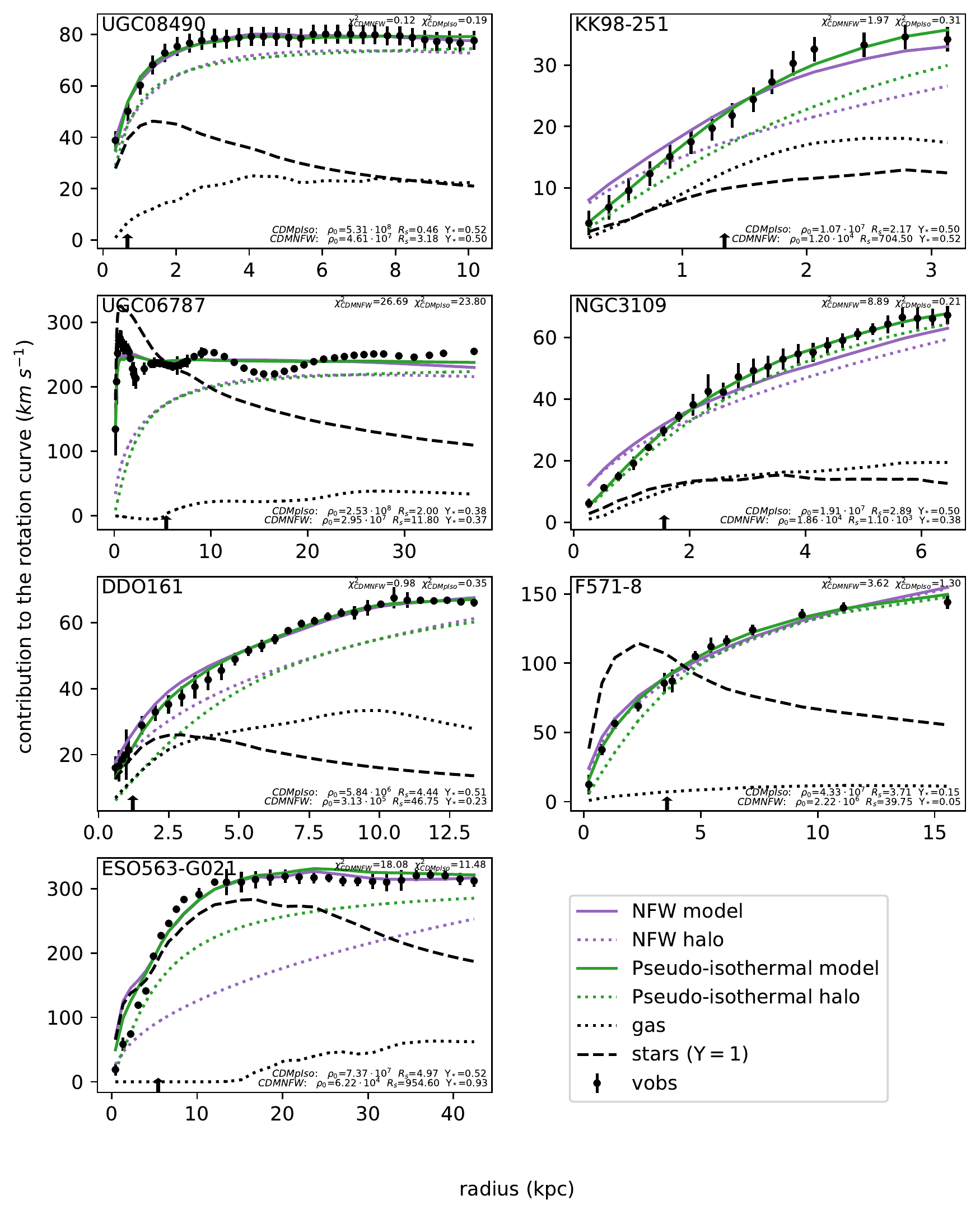}
\caption{ \label{fig:RChalo}Rotation curves fits of the CDM halo models. The solid lines "NFW model" and "Pseudo-isothermal model" are the total model velocity derived from the fits. The doted lines "NFW halo", "Pseudo-isothermal halo" are the contribution of the halo to the total velocity.}
\end{figure}

\begin{figure}[H]
\plotone{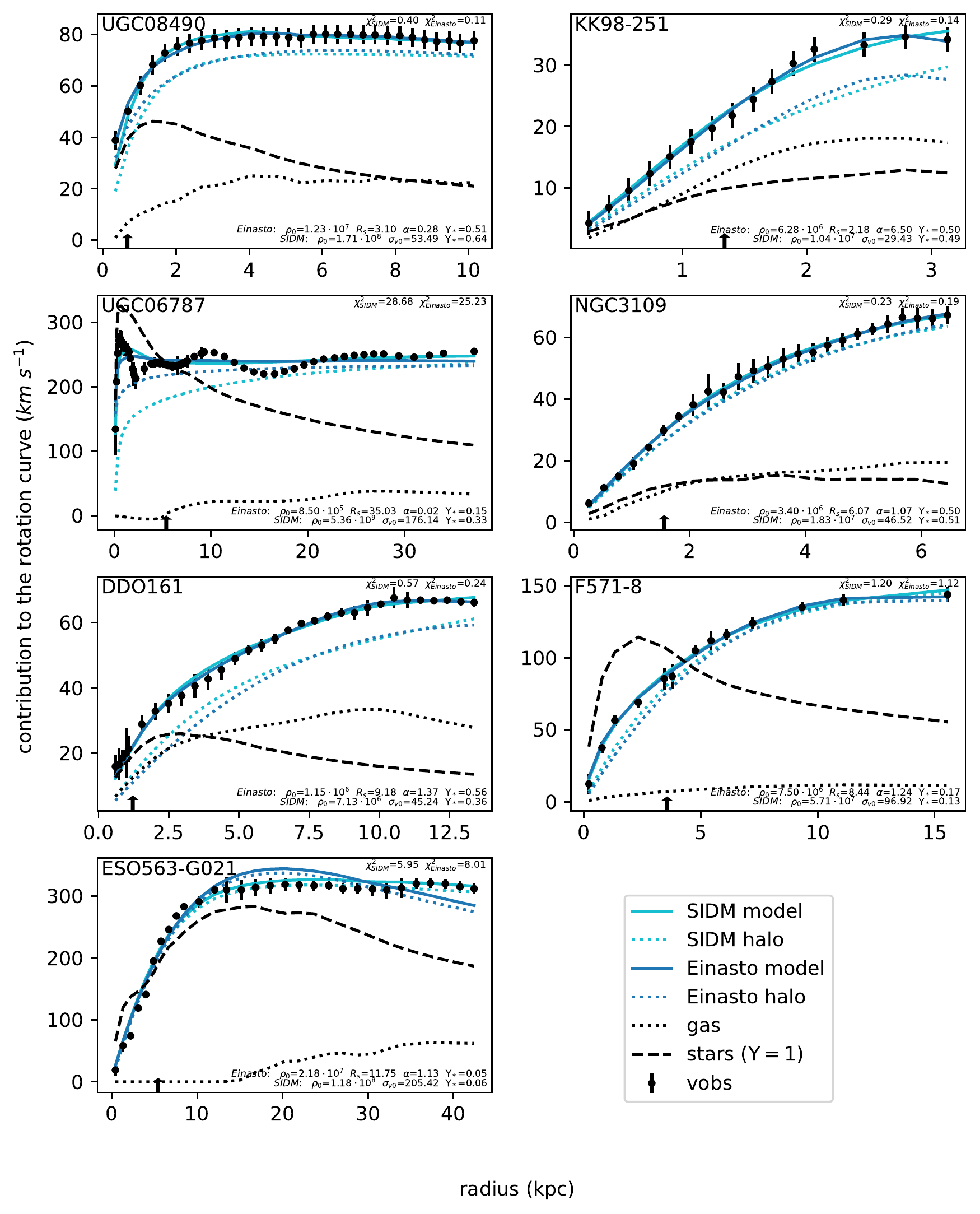}
\caption{ \label{fig:RCsidm}Rotation curves fits of the SIDM and Einasto models. The solid lines "SIDM model" and "Einasto model" are the total model velocity derived from the fits. The doted lines "SIDM halo" and "Einasto halo" are the contribution of the halo to the total velocity. }
\end{figure}

\begin{figure}[H]
\plotone{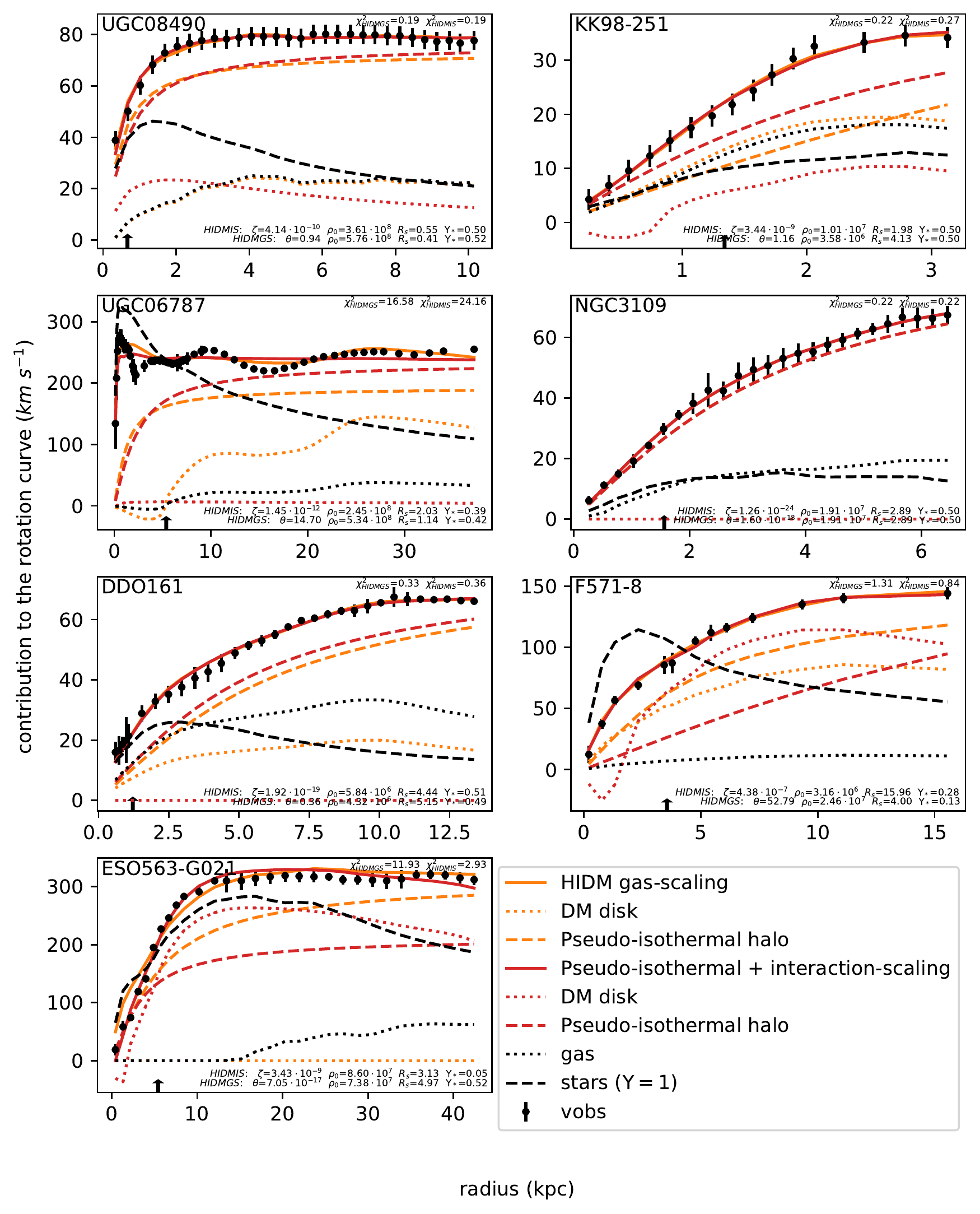}
\caption{ \label{fig:RCHIDM} Rotation curves fits of the HIDM models.  The solid lines are the total velocity derived from the model. The doted lines are the contribution of the DM disks to the total velocity. The dashed line is the contribution of the  halos to the total velocity.  }
\end{figure}

\begin{figure}[H]
\plotone{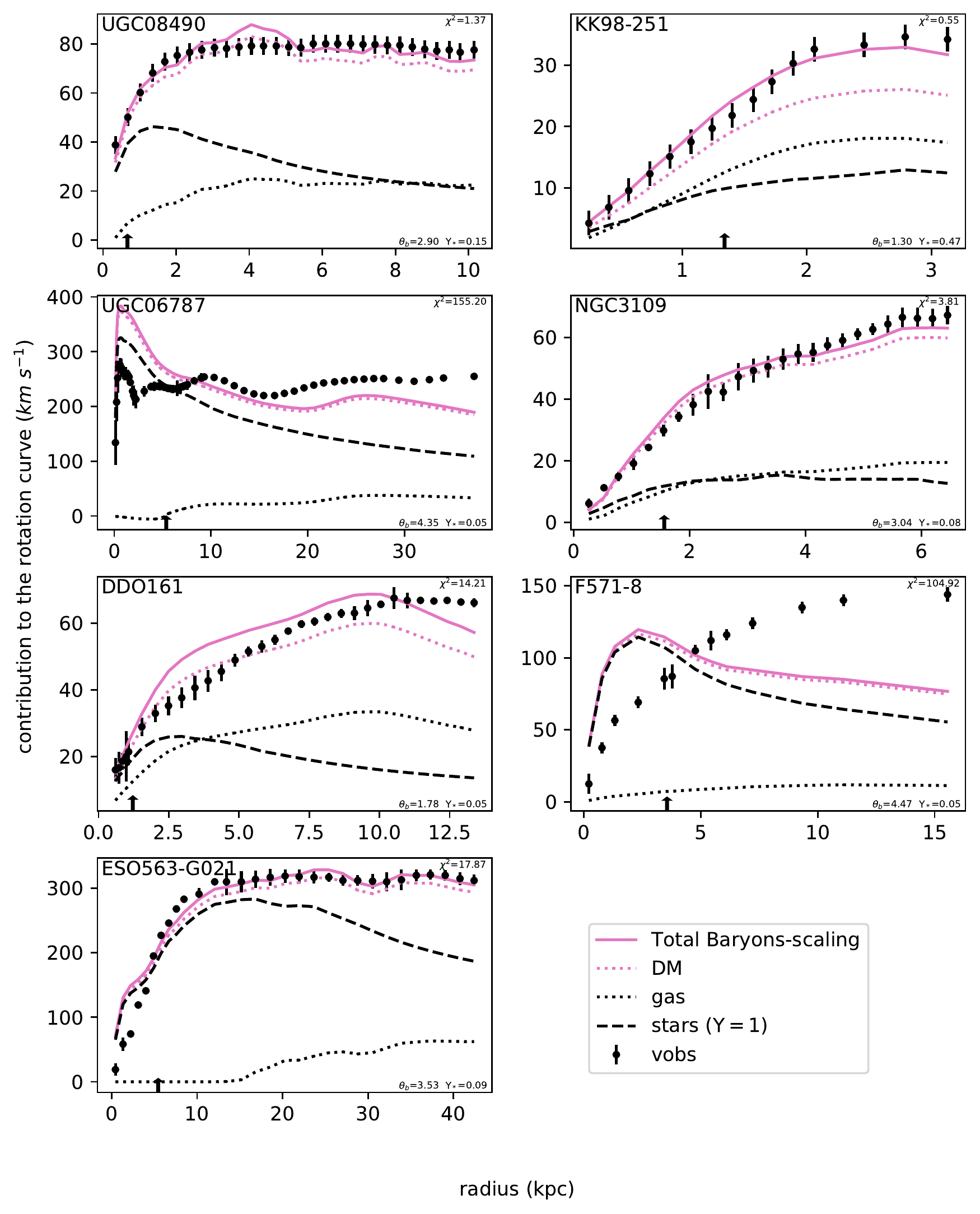}
\caption{ \label{fig:RCtotBS}Rotation curves fits of the total-baryon-scaling model. The green solid line is the total velocity derived from the baryon-scaling model. The green doted line is the contribution of the DM disk component to the total velocity. }
\end{figure}

\begin{figure}[H]
\plotone{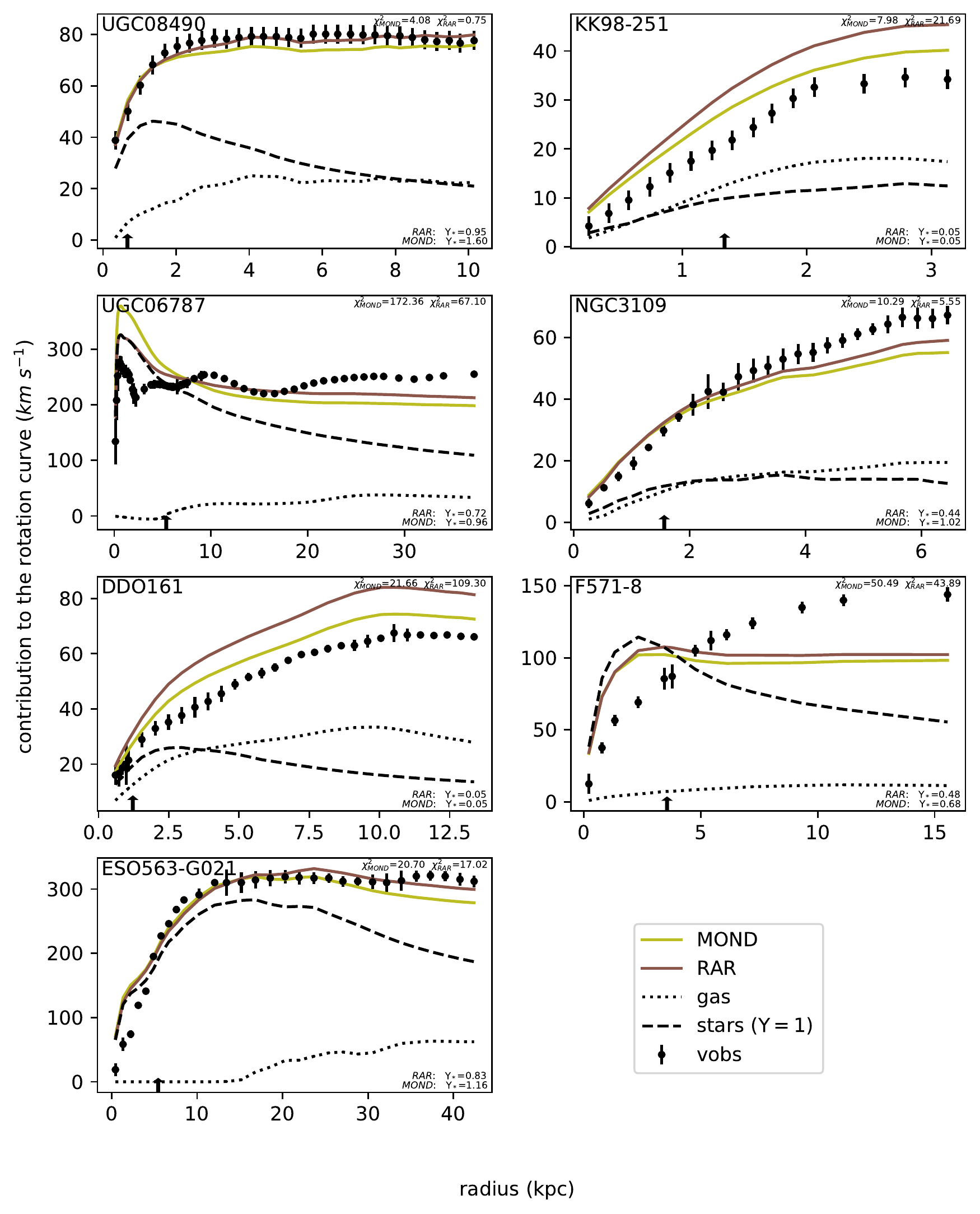}
\caption{ \label{fig:RCMOND} Rotation curves fits of MOND models. The green line is the total velocity derived from the historic MOND, the brown line is the is the total velocity from the RAR model. }
\end{figure}

\begin{figure*}[t]
\plotone{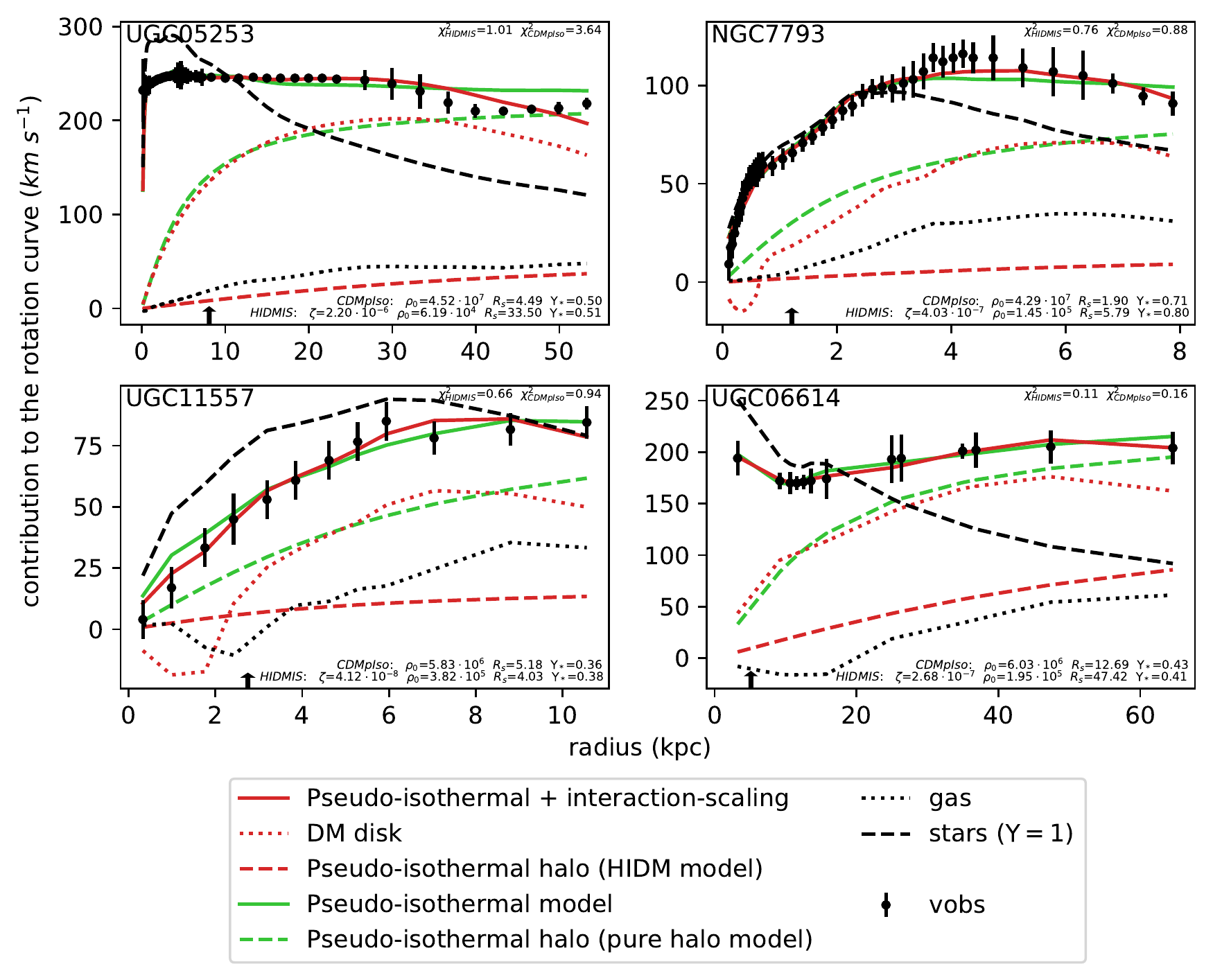}
\caption{HIDM interaction-scaling (red) and pseudo-isothermal (green) fits for the four galaxies that have a DM disk more massive than the DM halo. The contribution of the stars is displayed for $\Upsilon_*=1M_\odot/L_\odot$; the best fit value of $\Upsilon_*$, $\chi^2$ and other fit parameters for each galaxy are given in the panel legend. \label{fig:massiveDMdisk}}
\end{figure*}

To enable the reader to appreciate the successes and inadequacies of the different models for explaining the diversity of rotation curves, we show the rotation curve data for 7 illustrative galaxies along with the different models' best fits.  These 7 galaxies provide a good sample of the variety of fits from among the galaxies in the SPARC database whose rotation curves do not lack data near the galactic center.  Figure \ref{fig:RC7} allows all of the model predictions to be seen together, for each of the 7 galaxies.  Figures  \ref{fig:RChalo} - \ref{fig:RCMOND} give a more detailed view of how the different models achieve their best fit; in all cases the stellar contribution is shown for $\Upsilon_*=1M_\odot/L_\odot$.  Fig. \ref{fig:RChalo} and \ref{fig:RCsidm} display the fits of the four models with DM exclusively in halos:  NFW, pseudo-isothermal, Einasto and SIDM.  Fig. \ref{fig:RCHIDM} displays the fits of the two HIDM models having a pseudo-isothermal halo in addition to a DM disk derived in two different ways from the measured gas distribution, as detailed in Secs. \ref{sec:HIDMgs} and \ref{sec:HIDMis}.
Fig. \ref{fig:RCtotBS} shows fits with the total-baryon-scaling model and Fig. \ref{fig:RCMOND} shows the MOND and RAR model results. Finally, Fig. \ref{fig:massiveDMdisk} shows the RC fits for HIDM-IS, for four galaxies in which the DM disk is more massive than the DM halo.

The 7 galaxies we adopt for illustrative purposes are:
\begin{itemize}

\item UGC06787: The rotation curve contains oscillations that the baryonic components cannot explain alone. As these oscillations come from the gas contribution, the model that best fits this galaxy is the HIDM gas-scaling model. The very peaked inner star contribution and the rotation curve oscillations make the SIDM fits particularly bad. The gas contribution is too small to explain the oscillations in a SIDM context. 

\item NGC3109 is very well fitted by the pseudo-isothermal halo alone which gives results almost equal to the HIDM gas-scaling model. This is due to the fact that the contribution of the DM disk to the rotation curve is negligible.

\item KK98-251 is a small galaxy. MOND gives a poor description here because $v_{tot}$ is increasing at large radius while the baryonic components are flat.

\item UGC08490 is one of the galaxies with a characteristically flat rotation curve and slowly increasing inner curve for which MOND models gives particularly good results. The RAR MOND model (Equation \ref{eq:rar}) performs even better here. The smooth core and flat rotation curve make the SIDM model work particularly well, with $r_1=8.4\, {\rm kpc}$.

\item F571-8: MOND fails to explain this rotation curve. Indeed, the lack of gas contribution at large radius does not allows to explain the slowly increasing pattern of the rotation curve. Thus, the models which add a halo are particularly efficient compared to MOND. In addition, we note a very strong contribution due to the DM disk in the HIDM gas-scaling model.

\item ESO563-G021 is relatively well fitted by the total-baryons-scaling model. Here, the optimal fit for the HIDM gas-scaling model does not involve a DM disk. This example is typical of a subset of about 70 galaxies with little or no DM disk ($\theta < 10^{-4}$).

\item DDO161: the pseudo-isothermal halo alone gives a good fit. Here, the RAR MOND model gives worse results than the first MOND model.

\end{itemize}

\newpage

\newpage

\newpage

\newpage

\bibliography{bib}{}

\begin{thebibliography}{}
\expandafter\ifx\csname natexlab\endcsname\relax\def\natexlab#1{#1}\fi
\providecommand{\url}[1]{\href{#1}{#1}}
\providecommand{\dodoi}[1]{doi:~\href{http://doi.org/#1}{\nolinkurl{#1}}}
\providecommand{\doeprint}[1]{\href{http://ascl.net/#1}{\nolinkurl{http://ascl.net/#1}}}
\providecommand{\doarXiv}[1]{\href{https://arxiv.org/abs/#1}{\nolinkurl{https://arxiv.org/abs/#1}}}

\bibitem[{Blumenthal {et~al.}(1984)Blumenthal, Faber, Primack, \&
  Rees}]{Blumenthal1984}
Blumenthal, G.~R., Faber, S.~M., Primack, J.~R., \& Rees, M.~J. 1984, Nature,
  311, 517, \dodoi{10.1038/311517a0}

\bibitem[{{Bullock} \& {Boylan-Kolchin}(2017)}]{Bullock2017}
{Bullock}, J.~S., \& {Boylan-Kolchin}, M. 2017, \araa, 55, 343,
  \dodoi{10.1146/annurev-astro-091916-055313}

\bibitem[{Burkert(1995)}]{Burkert1996}
Burkert, A. 1995, The Astrophysical Journal, 447, \dodoi{10.1086/309560}

\bibitem[{{Casertano}(1983)}]{Casertano1983}
{Casertano}, S. 1983, \mnras, 203, 735, \dodoi{10.1093/mnras/203.3.735}

\bibitem[{Chan {et~al.}(2015)Chan, Kereš, Oñorbe, Hopkins, Muratov,
  Faucher-Giguère, \& Quataert}]{Chan2015}
Chan, T.~K., Kereš, D., Oñorbe, J., {et~al.} 2015, Monthly Notices of the
  Royal Astronomical Society, 454, 2981, \dodoi{10.1093/mnras/stv2165}

\bibitem[{Chemin {et~al.}(2011)Chemin, de~Blok, \& Mamon}]{Chemin2011}
Chemin, L., de~Blok, W. J.~G., \& Mamon, G.~A. 2011, The Astronomical Journal,
  142, 109, \dodoi{10.1088/0004-6256/142/4/109}

\bibitem[{Durkalec {et~al.}(2015)}]{Durkalec2014}
Durkalec, A., {et~al.} 2015, Astron. Astrophys., 576, L7,
  \dodoi{10.1051/0004-6361/201425532}

\bibitem[{Dvorkin {et~al.}(2014)Dvorkin, Blum, \&
  Kamionkowski}]{Dvorkin:2013cea}
Dvorkin, C., Blum, K., \& Kamionkowski, M. 2014, Phys. Rev. D, 89, 023519,
  \dodoi{10.1103/PhysRevD.89.023519}

\bibitem[{{Einasto}(1965)}]{Einasto1965}
{Einasto}, J. 1965, Trudy Astrofizicheskogo Instituta Alma-Ata, 5, 87.
\newblock \url{https://ui.adsabs.harvard.edu/abs/1965TrAlm...5...87E}

\bibitem[{Farrar(2017{\natexlab{a}})}]{Farrar2017eqq}
Farrar, G.~R. 2017{\natexlab{a}}.
\newblock \doarXiv{1708.08951}

\bibitem[{Farrar(2017{\natexlab{b}})}]{Farrar2017ysn}
Farrar, G.~R. 2017{\natexlab{b}}, in 35th International Cosmic Ray Conference,
  Proceedings of Science, Trieste, Italy, 929.
\newblock \doarXiv{1711.10971}

\bibitem[{Farrar(2018)}]{fDMtoB18}
---. 2018, A precision test of the nature of Dark Matter and a probe of the QCD
  phase transition.
\newblock \doarXiv{1805.03723}

\bibitem[{Farrar {et~al.}(2020)Farrar, Wang, \& Xu}]{Farrar2020}
Farrar, G.~R., Wang, Z., \& Xu, X. 2020, Dark Matter Particle in QCD.
\newblock \doarXiv{2007.10378}

\bibitem[{Gentile {et~al.}(2004)Gentile, Salucci, Klein, Vergani, \&
  Kalberla}]{Gentile2004}
Gentile, G., Salucci, P., Klein, U., Vergani, D., \& Kalberla, P. 2004, Monthly
  Notices of the Royal Astronomical Society, 351, 903,
  \dodoi{10.1111/j.1365-2966.2004.07836.x}

\bibitem[{Gluscevic \& Boddy(2018)}]{Gluscevic:2017ywp}
Gluscevic, V., \& Boddy, K.~K. 2018, Phys. Rev. Lett., 121, 081301,
  \dodoi{10.1103/PhysRevLett.121.081301}

\bibitem[{Hayashi {et~al.}(2007)Hayashi, Navarro, \& Springel}]{Hayashi2007}
Hayashi, E., Navarro, J.~F., \& Springel, V. 2007, Monthly Notices of the Royal
  Astronomical Society, 377, 50, \dodoi{10.1111/j.1365-2966.2007.11599.x}

\bibitem[{{Hui} {et~al.}(2017){Hui}, {Ostriker}, {Tremaine}, \&
  {Witten}}]{hotw17}
{Hui}, L., {Ostriker}, J.~P., {Tremaine}, S., \& {Witten}, E. 2017, \prd, 95,
  043541, \dodoi{10.1103/PhysRevD.95.043541}

\bibitem[{{Jimenez} {et~al.}(2003){Jimenez}, {Verde}, \& {Oh}}]{Jimenez2003}
{Jimenez}, R., {Verde}, L., \& {Oh}, S.~P. 2003, \mnras, 339, 243,
  \dodoi{10.1046/j.1365-8711.2003.06165.x}

\bibitem[{Kalberla \& Kerp(2009)}]{Kalberla2009}
Kalberla, P.~M., \& Kerp, J. 2009, Annual Review of Astronomy and Astrophysics,
  47, 27, \dodoi{10.1146/annurev-astro-082708-101823}

\bibitem[{Kaplinghat {et~al.}(2014)Kaplinghat, Keeley, Linden, \&
  Yu}]{Kaplinghat2014}
Kaplinghat, M., Keeley, R.~E., Linden, T., \& Yu, H.-B. 2014, Phys. Rev. Lett.,
  113, 021302, \dodoi{10.1103/PhysRevLett.113.021302}

\bibitem[{Kaplinghat {et~al.}(2016)Kaplinghat, Tulin, \& Yu}]{Kaplinghat2016}
Kaplinghat, M., Tulin, S., \& Yu, H.-B. 2016, Phys. Rev. Lett., 116, 041302,
  \dodoi{10.1103/PhysRevLett.116.041302}

\bibitem[{Katz {et~al.}(2016)Katz, Lelli, McGaugh, Di~Cintio, Brook, \&
  Schombert}]{Katz2016}
Katz, H., Lelli, F., McGaugh, S.~S., {et~al.} 2016, Monthly Notices of the
  Royal Astronomical Society, 466, 1648, \dodoi{10.1093/mnras/stw3101}

\bibitem[{{Lelli} {et~al.}(2016){Lelli}, {McGaugh}, \& {Schombert}}]{Lelli2016}
{Lelli}, F., {McGaugh}, S.~S., \& {Schombert}, J.~M. 2016, \aj, 152, 157,
  \dodoi{10.3847/0004-6256/152/6/157}

\bibitem[{Li {et~al.}(2020)Li, Lelli, McGaugh, \& Schombert}]{Li2020}
Li, P., Lelli, F., McGaugh, S., \& Schombert, J. 2020, The Astrophysical
  Journal Supplement Series, 247, 31, \dodoi{10.3847/1538-4365/ab700e}

\bibitem[{Loizeau \& Farrar(2021)}]{Loizeau2021}
Loizeau, N., \& Farrar, G.~R. 2021, Non-spherical dark matter structures
  detection.
\newblock \doarXiv{2106.14915}

\bibitem[{Markevitch {et~al.}(2004)Markevitch, Gonzalez, Clowe, Vikhlinin,
  Forman, Jones, Murray, \& Tucker}]{Markevitch2004}
Markevitch, M., Gonzalez, A.~H., Clowe, D., {et~al.} 2004, The Astrophysical
  Journal, 606, 819, \dodoi{10.1086/383178}

\bibitem[{McGaugh {et~al.}(2016)McGaugh, Lelli, \& Schombert}]{McGaugh2016}
McGaugh, S.~S., Lelli, F., \& Schombert, J.~M. 2016, Phys. Rev. Lett., 117,
  201101, \dodoi{10.1103/PhysRevLett.117.201101}

\bibitem[{{McGaugh} \& {Schombert}(2014)}]{McGaugh2014}
{McGaugh}, S.~S., \& {Schombert}, J.~M. 2014, \aj, 148, 77,
  \dodoi{10.1088/0004-6256/148/5/77}

\bibitem[{{Meidt} {et~al.}(2014){Meidt}, {Schinnerer}, {van de Ven},
  {Zaritsky}, {Peletier}, {Knapen}, {Sheth}, {Regan}, {Querejeta},
  {Mu{\~n}oz-Mateos}, {Kim}, {Hinz}, {Gil de Paz}, {Athanassoula}, {Bosma},
  {Buta}, {Cisternas}, {Ho}, {Holwerda}, {Skibba}, {Laurikainen}, {Salo},
  {Gadotti}, {Laine}, {Erroz-Ferrer}, {Comer{\'o}n}, {Men{\'e}ndez-Delmestre},
  {Seibert}, \& {Mizusawa}}]{Meidt2014}
{Meidt}, S.~E., {Schinnerer}, E., {van de Ven}, G., {et~al.} 2014, \apj, 788,
  144, \dodoi{10.1088/0004-637X/788/2/144}

\bibitem[{{Milgrom}(1983)}]{Milgrom1983}
{Milgrom}, M. 1983, \apj, 270, 365, \dodoi{10.1086/161130}

\bibitem[{Navarro {et~al.}(1996)Navarro, Eke, \& Frenk}]{Navarro1996}
Navarro, J.~F., Eke, V.~R., \& Frenk, C.~S. 1996, Monthly Notices of the Royal
  Astronomical Society, 283, L72, \dodoi{10.1093/mnras/283.3.L72}

\bibitem[{{Navarro} {et~al.}(1996){Navarro}, {Frenk}, \& {White}}]{NFW1996}
{Navarro}, J.~F., {Frenk}, C.~S., \& {White}, S.~D.~M. 1996, \apj, 462, 563,
  \dodoi{10.1086/177173}

\bibitem[{{Navarro} {et~al.}(1997){Navarro}, {Frenk}, \& {White}}]{NFW1997}
---. 1997, \apj, 490, 493, \dodoi{10.1086/304888}

\bibitem[{Navarro {et~al.}(2004)Navarro, Hayashi, Power, Jenkins, Frenk, White,
  Springel, Stadel, \& Quinn}]{Navarro2004}
Navarro, J.~F., Hayashi, E., Power, C., {et~al.} 2004, Monthly Notices of the
  Royal Astronomical Society, 349, 1039,
  \dodoi{10.1111/j.1365-2966.2004.07586.x}

\bibitem[{{Noordermeer}(2006)}]{Noordermeer2006}
{Noordermeer}, E. 2006, PhD thesis, Groningen: Rijksuniversiteit

\bibitem[{{Pontzen} \& {Governato}(2012)}]{Pontzen2012}
{Pontzen}, A., \& {Governato}, F. 2012, \mnras, 421, 3464,
  \dodoi{10.1111/j.1365-2966.2012.20571.x}

\bibitem[{Ren {et~al.}(2019)Ren, Kwa, Kaplinghat, \& Yu}]{Ren2019}
Ren, T., Kwa, A., Kaplinghat, M., \& Yu, H.-B. 2019, Phys. Rev. X, 9, 031020,
  \dodoi{10.1103/PhysRevX.9.031020}

\bibitem[{{Rubin} {et~al.}(1980){Rubin}, {Ford}, \& {Thonnard}}]{Rubin1980}
{Rubin}, V.~C., {Ford}, W.~K., J., \& {Thonnard}, N. 1980, \apj, 238, 471,
  \dodoi{10.1086/158003}

\bibitem[{Santos-Santos {et~al.}(2020)Santos-Santos, Navarro, Robertson,
  Benítez-Llambay, Oman, Lovell, Frenk, Ludlow, Fattahi, \& Ritz}]{Santos2020}
Santos-Santos, I. M.~E., Navarro, J.~F., Robertson, A., {et~al.} 2020, Monthly
  Notices of the Royal Astronomical Society, 495, 58,
  \dodoi{10.1093/mnras/staa1072}

\bibitem[{Scarpa(2006)}]{Scarpa2006}
Scarpa, R. 2006, AIP Conf. Proc., 822, 253, \dodoi{10.1063/1.2189141}

\bibitem[{{Schombert} \& {McGaugh}(2014)}]{Schombert2014}
{Schombert}, J., \& {McGaugh}, S. 2014, \pasa, 31, e036,
  \dodoi{10.1017/pasa.2014.32}

\bibitem[{Schombert {et~al.}(2018)Schombert, McGaugh, \& Lelli}]{Schombert2018}
Schombert, J., McGaugh, S., \& Lelli, F. 2018, Monthly Notices of the Royal
  Astronomical Society, 483, 1496, \dodoi{10.1093/mnras/sty3223}

\bibitem[{Schutz {et~al.}(2018)Schutz, Lin, Safdi, \& Wu}]{Schutz+thinDMdisk18}
Schutz, K., Lin, T., Safdi, B.~R., \& Wu, C.-L. 2018, Phys. Rev. Lett., 121,
  081101, \dodoi{10.1103/PhysRevLett.121.081101}

\bibitem[{Spergel \& Steinhardt(2000)}]{Spergel2000}
Spergel, D.~N., \& Steinhardt, P.~J. 2000, Phys. Rev. Lett., 84, 3760,
  \dodoi{10.1103/PhysRevLett.84.3760}

\bibitem[{{Swaters} {et~al.}(2014){Swaters}, {Bershady}, {Martinsson},
  {Westfall}, {Andersen}, \& {Verheijen}}]{Swaters2014}
{Swaters}, R.~A., {Bershady}, M.~A., {Martinsson}, T.~P.~K., {et~al.} 2014,
  \apjl, 797, L28, \dodoi{10.1088/2041-8205/797/2/L28}

\bibitem[{{Swaters} {et~al.}(2012){Swaters}, {Sancisi}, {van der Hulst}, \&
  {van Albada}}]{Swaters2012}
{Swaters}, R.~A., {Sancisi}, R., {van der Hulst}, J.~M., \& {van Albada}, T.~S.
  2012, \mnras, 425, 2299, \dodoi{10.1111/j.1365-2966.2012.21599.x}

\bibitem[{van~den Bosch \& Swaters(2001)}]{Bosch2001}
van~den Bosch, F.~C., \& Swaters, R.~A. 2001, Monthly Notices of the Royal
  Astronomical Society, 325, 1017, \dodoi{10.1046/j.1365-8711.2001.04456.x}

\bibitem[{Wadekar \& Farrar(2021)}]{Wadekar2021}
Wadekar, D., \& Farrar, G.~R. 2021, Phys. Rev. D, 103, 123028,
  \dodoi{10.1103/PhysRevD.103.123028}

\bibitem[{Xu {et~al.}(2018)Xu, Dvorkin, \& Chael}]{Xu:2018efh}
Xu, W.~L., Dvorkin, C., \& Chael, A. 2018, Phys. Rev. D, 97, 103530,
  \dodoi{10.1103/PhysRevD.97.103530}

\bibitem[{Xu \& Farrar(2021)}]{Xu2021}
Xu, X., \& Farrar, G.~R. 2021, Resonant Scattering between Dark Matter and
  Baryons: Revised Direct Detection and CMB Limits.
\newblock \doarXiv{2101.00142}

\end{thebibliography}
\bibliographystyle{aasjournal}

\end{document}